\documentclass[reprint,twocolumn,secnumarabic,amssymb, aps, prl]{revtex4-2}

\usepackage{graphicx}
\usepackage{xcolor}
\usepackage{gensymb}
\usepackage{times}
\usepackage{amsmath}
\usepackage{float}
\usepackage{placeins}
\usepackage[normalem]{ulem}
\usepackage{braket}
\usepackage{bm}
\usepackage[colorlinks,linkcolor=blue,citecolor=blue,urlcolor=blue,filecolor=black]{hyperref}

\usepackage{tikz}

\newcommand{\twofermion}{
  \begin{tikzpicture}[baseline=(base)]
    \node (base) at (0.7pt,-4.8pt) {};
    \fill (0,0) circle (1.7pt); 
    \fill (0.15,-0.15) circle (1.7pt); 
  \end{tikzpicture}
}
\newcommand{\nofermion}{\text{\O}}

\newcommand{\new}[1]{#1}
\newcommand{\deleted}[1]{}

\newcommand{\naturepar}[1]{\vspace{10pt}\noindent\textbf{#1}\\}

\newcommand{\abs}[1]{\lvert#1\rvert}

\begin{document}
 \title{Probing the Kitaev honeycomb model on a neutral-atom quantum computer}
\author{
Simon~J.~Evered$^{1,*}$, Marcin~Kalinowski$^{1,*}$, Alexandra~A.~Geim$^{1}$, Tom~Manovitz$^{1}$, Dolev~Bluvstein$^{1}$, Sophie~H.~Li$^{1}$, Nishad~Maskara$^{1}$, Hengyun~Zhou$^{1,2}$, Sepehr~Ebadi$^{1,3}$, Muqing~Xu$^{1}$, Joseph~Campo$^{2}$, Madelyn~Cain$^{1}$, Stefan~Ostermann$^{1}$, Susanne~F.~Yelin$^{1}$, Subir~Sachdev$^{1}$, Markus~Greiner$^{1}$, Vladan~Vuleti\'{c}$^{4}$, and Mikhail~D.~Lukin$^{1,\dagger}$}
\affiliation{$^1$Department~of~Physics,~Harvard~University,~Cambridge,~MA~02138,~USA \looseness=-1\\ 
$^2$QuEra Computing Inc., Boston, MA 02135, USA\\
$^3$Department~of~Physics,~Massachusetts~Institute~of~Technology,~Cambridge,~MA~02139,~USA\\ $^4$Department of Physics and Research Laboratory of Electronics, Massachusetts Institute of Technology, Cambridge, MA 02139, USA\\
$^*$These authors contributed equally to this work $^\dagger$Corresponding Author; E-mail: lukin@physics.harvard.edu
}

\begin{abstract}
Quantum simulations of many-body systems are among the most promising applications of quantum computers~\cite{feynman_simulating_1982}. In particular, models based on strongly-correlated fermions are central to our understanding of quantum chemistry and materials problems~\cite{clinton_towards_2024}, and can lead to exotic, topological phases of matter~\cite{kitaev_anyons_2006,fu_superconducting_2008}. However, due to the non-local nature of fermions, such models are challenging to simulate with qubit devices~\cite{jordan_ber_1928}. Here we realize a digital quantum simulation architecture for two-dimensional fermionic systems based on reconfigurable atom arrays~\cite{bluvstein_quantum_2022}. We utilize a fermion-to-qubit mapping based on Kitaev's model on a honeycomb lattice~\cite{kitaev_anyons_2006}, in which fermionic statistics are encoded using long-range entangled states~\cite{verstraete_mapping_2005}. We prepare these states efficiently using measurement~\cite{lu_measurement_2022} and feedforward~\cite{bluvstein_logical_2024}, realize subsequent fermionic evolution through Floquet engineering~\cite{abanin_effective_2017,else_prethermal_2017} with tunable entangling gates~\cite{evered_high-fidelity_2023} interspersed with atom rearrangement, and improve results with built-in error detection. Leveraging this fermion description of the Kitaev spin model, we efficiently prepare topological states across its complex phase diagram~\cite{kalinowski_non-abelian_2023} and verify the non-Abelian spin liquid phase~\cite{kitaev_anyons_2006} by evaluating an odd Chern number~\cite{berry_quantal_1997,alicea_new_2012}. We further explore this two-dimensional fermion system by realizing tunable dynamics and directly probing fermion exchange statistics. Finally, we simulate strong interactions and study dynamics of the Fermi-Hubbard model on a square lattice. These results pave the way for digital quantum simulations of complex fermionic systems for materials science, chemistry~\cite{maskara_programmable_2023}, and high-energy physics~\cite{maldacena2023simplequantumdescribesblack}.
\end{abstract}
\maketitle

Quantum computers have the potential to fundamentally advance our ability to simulate strongly correlated many-body quantum systems, which can be challenging to simulate classically~\cite{feynman_simulating_1982}. Fermionic models are especially important due to their central role in understanding phenomena across chemistry, materials science, and fundamental physics~\cite{altman_quantum_2021}. Recently, analog quantum simulators~\cite{gross_quantum_sim_review} based on fermionic atoms in optical lattices have been advancing our understanding of the Fermi-Hubbard model~\cite{Xu2023,Shao2024,chalopin_probing_2024}. However, digital quantum simulation approaches offer new possibilities with improved control and programmability, providing access to a much wider range of models and, eventually, to fault-tolerant operation. Early experiments demonstrated key building blocks~\cite{Barends2015, Arute2020, mi_noise-resilient_2022} of digital fermionic encodings, and recent studies have extended the work to two dimensions~\cite{cochran_visualizing_2024, nigmatullin_experimental_2024,hemery_measuring_2024}.

Implementing fermionic models on gate-based quantum computers presents unique challenges associated with encoding the non-local nature of fermions within qubit systems. As a result, fermion-to-qubit encodings require either macroscopic operators to represent local fermion hopping~\cite{jordan_ber_1928} or rely on long-range entangled ancilla states to preserve the locality of operations~\cite{verstraete_mapping_2005,derby_compact_2021, chen_equivalence_2023}. The latter case requires preserving topological order throughout computations to accurately represent fermionic particles, and can additionally be related to models that host exotic states of matter with topological order~\cite{verstraete_mapping_2005}. A paradigmatic case is Kitaev’s honeycomb model~\cite{kitaev_anyons_2006}, which features a non-Abelian chiral spin liquid with a topologically non-trivial energy band. While certain Abelian and non-Abelian states have recently been prepared in analog and digital systems~\cite{semeghini_probing_2021,satzinger_realizing_2021, andersen_non-abelian_2023,iqbal_non-abelian_2024}, the Kitaev model and its extensions~\cite{Takagi.2019, duan_opticallattice_2003} constitute a promising platform for exploring both static and dynamic properties of such exotic phases of matter\new{~\cite{google_kitaev_paper}}, including the dynamical interplay between topological order and strongly correlated fermions.

Here, we realize digital quantum simulations of Kitaev's honeycomb model using a reconfigurable atom array processor. Leveraging measurement-based topological state preparation, tunable Floquet circuits, and error detection, we probe the phase diagram and dynamics of this model. In particular, we prepare and characterize several low-energy states, including that of the non-Abelian spin liquid phase. We further utilize this model as an efficient fermion-to-qubit encoding, investigating fermion properties through quench dynamics and engineering strong interactions to realize the Fermi-Hubbard model on a square lattice.

\begin{figure*}
\includegraphics[width=18.0cm]{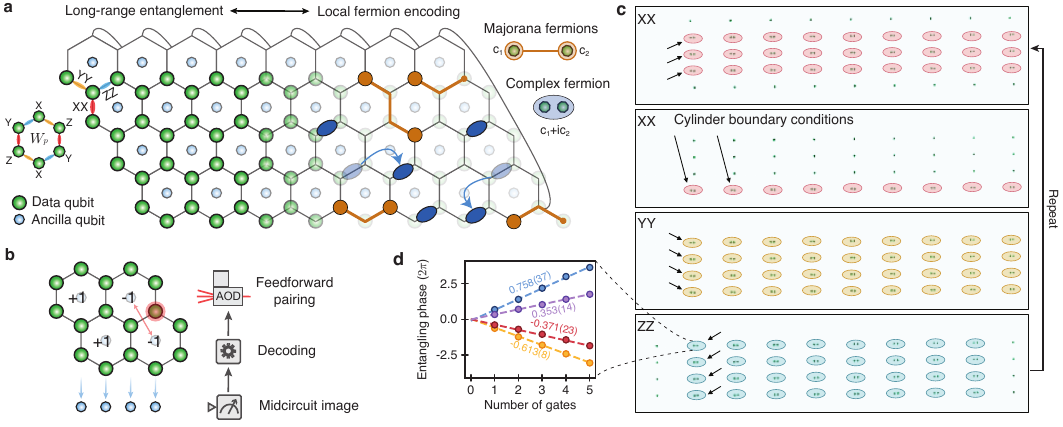}
\caption{\textbf{Digital quantum simulations with reconfigurable atom arrays.} \textbf{a,} The honeycomb lattice 
used in this work with 104 total atomic qubits and periodic boundary conditions along the shorter direction, forming a cylinder (see also Extended Data Fig.~\ref{fig:ED_ExperimentSequence}a). The qubits are encoded in $^{87}$Rb atoms and entangling gates are realized through excitation to interacting Rydberg states. To encode fermion statistics, 
we prepare a long-range entangled state characterized by hexagonal plaquette operators $W_p{=}{\rm X}_1{\rm Z}_2{\rm Y}_3{\rm X}_4{\rm Z}_5{\rm Y}_6$ that commute with the unitary evolution and are therefore conserved. The encoded Majorana fermions live on vertices, at the ends of operator strings, and conventional (complex) fermions are formed by combining two Majoranas along a chosen link orientation. \textbf{b,} The long-range entangled state is prepared using mid-circuit measurement of ancilla qubits, and plaquettes are deterministically flipped to be +1 using conditional single-qubit gates (red circle). \textbf{c,} The Floquet evolution cycle consists of atom reconfiguration interspersed with tunable entangling gates and global basis changes. \textbf{d,} Accumulated entangling phase upon repeated application of $\exp(i\pi\theta {\rm [Z{\otimes}Z]/4})$ gates with a variable angle $\theta$ realized through fast, parameterized laser pulses (Extended Data Fig.~\ref{fig:ED_EntanglingGates}). The values show extracted values of $\theta$, and error bars represent one standard deviation.}
\label{fig:fig1}
\end{figure*}

\naturepar{Hardware-efficient digital quantum simulations}
The key idea behind our approach is the use of long-range entanglement to encode fermionic statistics~\cite{verstraete_mapping_2005}. In particular, while conventional fermion encodings~\cite{jordan_ber_1928} require high-weight qubit operators, topologically ordered states contain long-range correlations that can natively encode these properties. The Kitaev model is a paradigmatic example of such a fermion encoding based on topological order~\cite{kitaev_anyons_2006}. Here qubits are located on the vertices of the honeycomb lattice (Fig.~\ref{fig:fig1}a and Extended Data Fig.~\ref{fig:ED_ExperimentSequence}a) and are effectively coupled through anisotropic two-qubit interactions, 
\begin{align}
     K_{ij}^{\rm X} &= -J_{\rm X} {\rm X}_i{\rm X}_j, & K_{ij}^{\rm Y} &= -J_{\rm Y} {\rm Y}_i {\rm Y}_j, &K_{ij}^Z &= -J_{\rm Z} {\rm Z}_i {\rm Z}_j,\label{eq:spinint}
\end{align}
acting along the three distinct link directions with interaction strengths $J_{\rm X}$, $J_{\rm Y}$ and $J_{\rm Z}$, as defined in Fig.~\ref{fig:fig1}a. The X$_i$,Y$_i$,Z$_i$ are the Pauli operators acting on qubit $i$, and we refer to the three link orientations as XX, YY, and ZZ links, respectively. In such a system, there are long-range entangled states which correspond to the common eigenstates of the conserved, mutually commuting plaquette operators, $W_p{=}{\rm X}_1{\rm Z}_2{\rm Y}_3{\rm X}_4{\rm Z}_5{\rm Y}_6$
(Fig.~\ref{fig:fig1}a). 
When acting on such states, the spin interactions in Eq.~\eqref{eq:spinint} can be described as Majorana hopping operators, $ic_ic_j$, along appropriate links (Extended Data Fig.~\ref{fig:ED_FermionEncoding}). The resulting system of Majorana fermions, $c_i$, located on the vertices of the lattice, realizes the fermionic mapping of Kitaev's model~\cite{kitaev_anyons_2006}.

For the experimental realization of this model, we utilize a neutral-atom quantum computer based on a dynamically reconfigurable atom array composed of 72 data qubits and 32 ancilla qubits (Fig.~\ref{fig:fig1}). Using the apparatus described previously in Refs.~\cite{ebadi_quantum_2021,bluvstein_quantum_2022,evered_high-fidelity_2023,bluvstein_logical_2024}, qubits are encoded in the long-lived hyperfine levels of $^{87}$Rb atoms, individually trapped in optical tweezers, and can be reconfigured during computation while maintaining qubit coherence~\cite{bluvstein_quantum_2022}. Fully programmable global and local single-qubit gates are implemented via fast Raman transitions~\cite{bluvstein_logical_2024}, while high-fidelity entangling gates are achieved by exciting atoms to interacting Rydberg states~\cite{evered_high-fidelity_2023}. A key new element involves the parallel implementation of a family of gates, ${\rm ZZ}(\theta){=}\exp(i\pi\theta {\rm [Z{\otimes}Z]/4})$, with tunable angle $\theta$ (Fig.~\ref{fig:fig1}d, Extended Data Fig.~\ref{fig:ED_EntanglingGates}). We intersperse these gates with atom motion and single-qubit rotations to realize Floquet circuits (Fig.~\ref{fig:fig1}c), which consist of repeated digital evolution under the two-qubit 
interactions in Eq.~\eqref{eq:spinint} applied along appropriate links.

To efficiently create the topological order~\cite{lu_measurement_2022}, ancilla qubits are used to project data qubits into the target long-range entangled state with cylindrical boundary conditions (Fig.~\ref{fig:fig1}b). This is achieved by measuring commuting plaquette operators $W_p$ in two steps (Fig.~\ref{fig:fig2}a). First, ancilla qubits are used to measure weight-4 operators on one sublattice of the data qubits, initially all prepared in $\ket{0}$. Such a mid-circuit measurement projects the system into a toric code state~\cite{kitaev_fault-tolerant_2003}. In the second step, the two sublattices are entangled with parallel controlled-Y gates, completing the weight-6 plaquette operators required for this fermion encoding (Extended Data Fig.~\ref{fig:ED_AncillaMeasurement}). Although the ancilla measurement outcomes are random $\pm$1 values, we apply conditional (feedforward) local single-qubit Z gates to deterministically prepare all plaquettes to have +1 value in the absence of errors (Fig.~\ref{fig:fig2}a and Extended Data Fig.~\ref{fig:ED_FeedforwardMethods}a).

To characterize the resulting state, we measure the plaquette operators $W_p$, finding an average parity of 0.444(4) (Fig.~\ref{fig:fig2}b). Furthermore, we measure longer closed-loop operators, enclosing up to four hexagons, and observe finite parity in all cases (Fig.~\ref{fig:fig2}\deleted{d}\new{c} and Extended Data Fig.~\ref{fig:ED_FeedforwardMethods}c). We can improve the quality of this topological order by performing error detection, by noting that for the cylindrical geometry used here, the product of ancilla values in a given column must have even parity in the absence of errors (Extended Data Fig.~\ref{fig:ED_FeedforwardMethods}b). Using this decoding postselection method (Fig.~\ref{fig:fig2}\deleted{c}\new{d}), we observe an improvement in the plaquette parity to 0.57(1).

\naturepar{Exploring the Kitaev model}
The highly anisotropic interaction structure of the Kitaev model, Eq.~\eqref{eq:spinint}, gives rise to a rich phase diagram, featuring several distinct topological phases (Fig.~\ref{fig:fig3}a). Of particular interest is the non-Abelian spin liquid phase, known as phase B, which emerges when the three coupling strengths $J_{\rm X}$,$J_{\rm Y}$ and $J_{\rm Z}$ are sufficiently uniform~\cite{kitaev_anyons_2006}. The fermionic description of the model results in its low-energy states being characterized by the underlying spin-liquid order and effective fermion correlations~\cite{kitaev_anyons_2006}. The initial state after feedforward corresponds to a fixed-point state A$_{\rm Z}^{\rm I}$ of the Abelian A$_{\rm Z}$ phase ($J_{\rm X/Y}{=}0$), where all ZZ-link operators should be +1. We observe a mean ZZ-link parity of 0.883(8), and the measured plaquette operators characterize the spin-liquid nature of the prepared state (Fig.~\ref{fig:fig2}b).

To evolve from this initial state (A$_{\rm Z}^{\rm I}$) to target states of interest in the phase diagram, we apply a numerically optimized version of a Floquet unitary circuit~\cite{kalinowski_non-abelian_2023} constructed from repeated application of the nearest-neighbor gates as illustrated in Fig.~\ref{fig:fig1}c, with optimized entangling phases. We find that a circuit depth of 6, as shown in Fig.~\ref{fig:fig3}b, is sufficient \new{at this system size} to probe all points of interest on the phase diagram ({\new{as characterized by Majorana correlations, see} Methods), consistent with the slow gap closing with system size~\cite{kalinowski_non-abelian_2023}. We measure the plaquette parity values to verify the preservation of the topological order and observe that they decay with increasing circuit depth (Fig.~\ref{fig:fig3}b), as expected due to accumulating incoherent errors. To mitigate this, we utilize a new type of state-selective readout, which converts atomic state into spatial position~\cite{wu_stern-gerlach_2019, other_aa_paper}, allowing us to differentiate between lost atoms and the two computational states~\cite{scholl_erasure_2023,ma_high-fidelity_2023}. By detecting atom loss and employing the error-detecting technique based on ancilla results, we observe that we can better preserve the plaquette expectation values during computation (Fig.~\ref{fig:fig3}b).

\begin{figure}
\includegraphics[width=89mm]{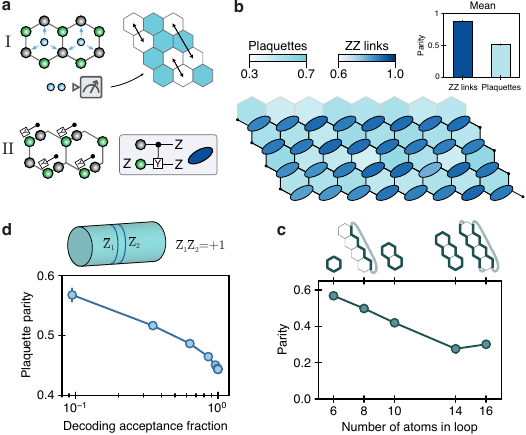}
\caption{\textbf{Measurement-based preparation of topological order.} \textbf{a,} A long-range entangled state of the toric-code type is prepared using a depth-3 circuit, independent of the system size, accompanied by midcircuit readout of the ancilla qubits (I). The measurement results are random up to parity constraints along the periodic direction. A feedforward step realized through FPGA-triggered single-qubit Z rotations pairs the -1 outcomes (white hexagons). Finally, a parallel controlled-Y operation creates the weight-6 plaquettes (hexagons) and initializes the ZZ-link operators (ovals) to all be +1 (II). \textbf{b,} Expectation values of the weight-6 plaquettes and weight-2 ZZ-links across the array of 72 data and 32 ancilla qubits with cylindrical boundary conditions. The inset shows values averaged across the system. \textbf{c,} Parity expectation values of increasingly large loops, including loops that enclose 1, 2, 3, and 4 hexagons within a column and the loop around the cylinder (plotted for maximum decoding postselection, see Extended Data Fig.~\ref{fig:ED_FeedforwardMethods}c). The largest operator is equivalent to the product of two loops enclosing the cylinder. Error bars represent 68\% confidence intervals. \textbf{d,} The product of plaquette operators in each column is equivalent to loops enclosing the cylinder, which are +1 in the absence of errors. Postselection based on the parity of the measured ancilla values within each column improves the plaquette expectation values and the quality of the resulting fermion encoding.} 
\label{fig:fig2}
\end{figure}

To characterize states across the phase diagram, we measure two-point Majorana fermion correlations, $\braket{c_ic_j}$, which are constructed as products of the link operators along the path connecting the two sites (Fig.~\ref{fig:fig3}d). Therefore, all fermion correlations can be measured as open Pauli strings\new{, enabled by the underlying topological order}. 
Fig.~\ref{fig:fig3}c shows the measured expectation values of strings up to length 6, for three points in the phase diagram: the initial state after feedforward state preparation A$_{\rm Z}^{\rm I}$, another point in the Abelian phase A$_{\rm Z}^{\rm II}$, and a point in the non-Abelian phase B. We observe a buildup of string correlations as the system is moved towards phase B and find good agreement with theory predictions, under a noise model consisting of single-qubit Pauli errors and atom loss (Methods), across all prepared states.

We probe the nature of phase B by extracting the Chern number---a topological invariant that characterizes the curvature of the system's energy bands~\cite{berry_quantal_1997,aidelsburger_measuring_2015,cooper_topological_2019}. Noting that the prepared state should correspond to a low-energy state of a free-fermion Hamiltonian, we infer the momentum-space parent Hamiltonian ${H}_{\textbf{k}}$ from the measured real-space correlations of the state (Fig.~\ref{fig:fig3}e, Extended Data Fig.~\ref{fig:ED_MajoranaChern}c). Because the unit cell of the honeycomb lattice consists of two sites, the system possesses two Bloch bands and their Chern number can be numerically evaluated from the learned Hamiltonian (Extended Data Fig.~\ref{fig:ED_MajoranaChern}e). As the system transitions from the Abelian to the non-Abelian phase, we observe a change from a zero ($\rm{C}$=0) to a non-zero ($\rm{C}$=1) Chern number of the lowest-energy band (Fig.~\ref{fig:fig3}f). In phase B, an odd Chern number, combined with the underlying topological order (Fig.~\ref{fig:fig2}b \new{and Fig.~\ref{fig:fig3}b}), provides strong evidence of the phase's non-Abelian character (Methods). In particular, it guarantees that, in the presence of a flux, a non-Abelian, unpaired Majorana mode must be present in the spectrum at zero energy~\cite{kitaev_anyons_2006,alicea_new_2012}.

\begin{figure}[!t]
\centering
\includegraphics[width=89mm]{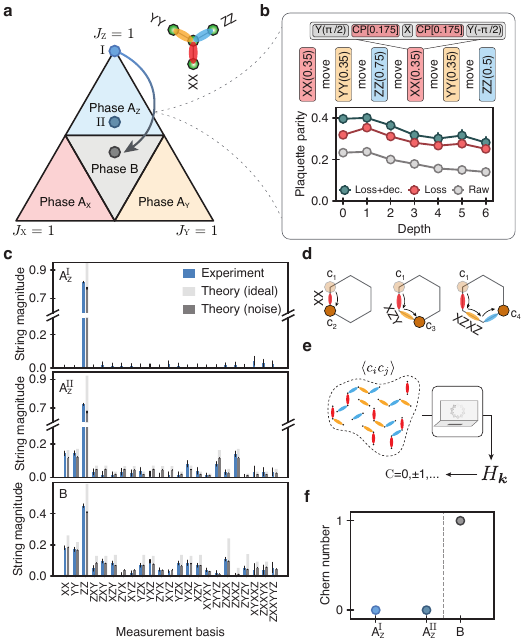}
\caption{\textbf{Kitaev model on a honeycomb lattice.} \textbf{a,} The honeycomb model, consisting of anisotropic spin interactions along the three directions of the honeycomb lattice ($J_{\rm X}{+}J_{\rm Y}{+}J_{\rm Z}{=}1$), exhibits topological order with three Abelian (A) phases and a single non-Abelian phase B. The toric code state in Fig.~\ref{fig:fig2} is the fixed-point state A$_{\rm Z}^{\rm I}$ of the Abelian A$_{\rm Z}$ phase ($J_{\rm Z}{=}1$). \textbf{b,} Starting from this initial state, we prepare different states on the phase diagram. Top: numerically optimized sequence of two-qubit gates used to prepare the non-Abelian phase B. Each circuit layer includes CPHASE (CP) gates and global single-qubit rotations. Bottom: plaquette parity during the evolution, with postselection based on atom loss and decoding. \new{The plaquette parity is lower than in Fig.~\ref{fig:fig2} due to a longer sequence of gates and atom movements (Methods).} \textbf{c,} Pauli strings of different lengths measured on the three studied states and averaged over the bulk of the system (Extended Data Fig.~\ref{fig:ED_MajoranaChern}a-b). Error bars represent 68\% confidence intervals. \textbf{d,} In the fermion representation, the link operators are proportional to nearest-neighbor Majorana hoppings, $K^{{\rm X/Y/Z}}_{ij}{\propto} ic_i c_j$. Longer Pauli strings constructed from their products result in longer-range hopping operators. \textbf{e,} The free-fermion parent Hamiltonian ${H}_{\textbf{k}}$ can be reconstructed from measured two-point Majorana correlations. \textbf{f,} The Chern number $\rm{C}$ is evaluated using the learned parent Hamiltonians of the string distributions in \textbf{c}, resulting in $\rm{C}$=0 in phase A and $\rm{C}$=1 in phase B; the robustness of this procedure is explored in Methods (Extended Data Fig.~\ref{fig:ED_MajoranaChern}f-h). The non-Abelian phase B is characterized by the underlying topological order and an odd Chern number~\cite{kitaev_anyons_2006}.
}
\label{fig:fig3}
\end{figure}

\naturepar{Floquet evolution of fermion systems}
Having explored the topological properties in the Kitaev phase diagram, we now leverage them to simulate quench dynamics~\cite{de_leseleuc_observation_2019,spar_realization_2022,young_tweezer-programmable_2022} of fermionic particles. We focus on conventional fermions characterized by complex-valued operators, which can be realized by pairing two Majorana fermions, $a_i{=}(c_i{+}i c_{i'})/2$. In our realization, we pair them along the ZZ links, such that the resulting complex fermions form a square-lattice geometry (Methods), and the ${\rm Z}_i {\rm Z}_{i'}$ operator along an ($i$,$i'$) link is related to local fermion density, $n_{i} {=}(1{-}{\rm Z}_i{\rm Z}_{i'})/2$, with ${\rm Z}_i {\rm Z}_{i'}{=}$+1 denoting particle absence, $n_i{=}0$. In this framework, the ZZ-link operators in Fig.~\ref{fig:fig2}b characterize the initial vacuum state, corresponding to a background fermion density of $5.9(5)\%$.
\begin{figure*}
\centering
\includegraphics[width=180mm]{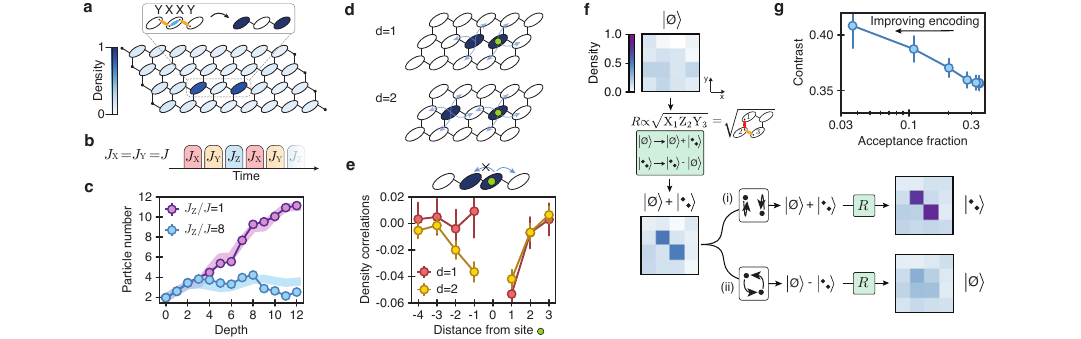}
\caption{\textbf{Probing fermion dynamics and statistics.} \textbf{a,} System initialized in a two-particle state with complex fermions defined along the ZZ links. An open Pauli string is used to create a fermion pair out of the initial vacuum state. \textbf{b,} Digital evolution under two-qubit interactions~in Eq.~\eqref{eq:spinint}, realizing Majorana fermion hopping between connected vertices. The hopping strength along the XX and YY links is fixed, $J_{\rm X}{=}J_{\rm Y}{=}J$, while the value of $J_{\rm Z}/J$ is tuned. \textbf{c,} Total number of complex fermions in the system during a quench under two different Floquet drives, for $J_{\rm Z}/J{=}1$ and $J_{\rm Z}/J{=}8$. The $K^{\rm Z}_{ij}$ term corresponds to the fermion particle number and increasing it enforces particle number conservation. The initial background fermion number resulting from state preparation errors is subtracted for clarity and the solid lines are the result of  noisy numerical simulations (Methods). \textbf{d,} System initialized with two fermions at distances d=1 or d=2 is evolved under the Floquet Hamiltonian with $J_{\rm Z}/J{=}8$. \textbf{e,} A horizontal cut of the two-point density-density correlation relative to the marked site (green dot) at depth-11. The directional asymmetry in hopping is consistent with Pauli exclusion, which prevents fermions from occupying the same site. \textbf{f,} Direct measurement of the fermion exchange phase. A superposition of vacuum and fermion-pair states is created with a sequence of local entangling gates, which realizes a partial creation unitary $R$ (Methods). The system is  evolved under two different particle-hopping protocols: (i) hopping to an intermediate site and hopping back and (ii) exchanging the two fermions. The evolution is followed by a second application of $R$, which maps the exchange phase to fermion presence. Color plots show the mean fermion density at each site in a 4x4 region embedded in the full lattice (Extended Data Fig.~\ref{fig:ED_Exchange}). 
\textbf{g,} Contrast between protocols (i) and (ii) as a function of decoding postselection. Error bars represent 68\% confidence intervals.
}
\label{fig:fig4}
\end{figure*}

To initialize real-space configurations of complex fermions, we start from the state A$_{\rm Z}^{\rm I}$ and create a fermion pair by applying an open Pauli string (e.g YXXY in Fig.~\ref{fig:fig4}a) that commutes with the plaquette operators but flips the desired ${\rm Z}_i {\rm Z}_{i'}$ operators such that $n_i{=}1$. Individual complex fermions can, in principle, be created by connecting such strings to the open boundary.

The dynamics of complex fermions is inherited from that of the Majorana operators, where the Kitaev interactions in Eq.~\eqref{eq:spinint} correspond to nearest-neighbor Majorana hopping terms $c_i c_j$ (Extended Data Fig.~\ref{fig:ED_FermionEncoding}). We first  study the dynamics of the two localized fermions under two different depth-12 Floquet circuits (Fig.~\ref{fig:fig4}b), focusing on the role of anisotropy between the couplings. Setting two directions to be equal, $J_{\rm X}{=}J_{\rm Y}{=}J$, and tuning the strength of the third direction $J_{\rm Z}$, we find that for $J_{\rm Z}{=}J$ the number of particles grows quickly, because the Majorana hopping terms do not naturally preserve the complex-fermion particle number (Fig.~\ref{fig:fig4}c and Extended Data Fig.~\ref{fig:ED_FermionHopping}a). By contrast, for $J_{\rm Z}{=}8J$, background particle creation is suppressed and the total particle number returns close to the initial value for a full effective Floquet cycle (around depths 6 and 12). This emergent particle conservation is the result of strong $K^{\rm Z}_{ij}$ terms, which corresponds to the total particle number, projecting the dynamics into approximate particle-conserving evolution~\cite{abanin_effective_2017,else_prethermal_2017}. Indeed, unitary dynamics result in an effective Floquet Hamiltonian closely resembling that of complex fermions, and the intermediate rise in particle number can be interpreted as Floquet micromotion (Methods, Extended Data Fig.~\ref{fig:ED_FermionHopping}b).

Focusing on the particle number conserving ($J_{\rm Z}{=}8J$) case, we evolve two initial states (Fig.~\ref{fig:fig4}d), with a complex fermion pair either one (d=1) or two (d=2) sites apart, and investigate the final fermion configuration at depth 11. We probe their hopping by measuring density-density correlations $G_{ij}{=}\langle n_i n_j\rangle{-}\langle n_i \rangle\langle n_j \rangle$. Such correlations capture hopping since a transported particle should be anticorrelated with its initial position (Methods). For free-fermion states, the $G_{ij}$ correlation is equal to the (negative) magnitude of the hopping operator between sites $i$ and $j$ (Methods). In Fig.~\ref{fig:fig4}e, we explore these correlations with respect to one of the initially occupied sites, normalized by its density. For the d=1 initial separation, we observe a strong asymmetry in the hopping direction, owing to the fermions hopping away from each other, consistent with Pauli exclusion. As expected, the asymmetry is greatly reduced for the d=2 separation.

\naturepar{Fermion exchange}
A defining characteristic of fermionic particles is their non-trivial exchange statistics, with the many-body wavefunction being antisymmetric under particle exchange. We directly probe the exchange statistics of the complex fermions through deterministic hopping, realized by a gate sequence designed to preserve particle number (Extended Data Figs.~\ref{fig:ED_FermionEncoding}~and~\ref{fig:ED_Exchange}). Because the exchange statistics manifest as a global phase, we perform a Ramsey-type experiment (Fig.~\ref{fig:fig4}f) in which we first apply a partial pair-creation operator $R{\propto}\sqrt{\rm X{\otimes} Z{\otimes}Y}$ to prepare a superposition state, $(\ket{\nofermion}{+}\ket{\twofermion})/\sqrt{2}$, where $\ket{\nofermion}$ is the initial vacuum state and $\ket{\twofermion}$ is the two-fermion state (Methods). Then, we compare two different fermion movement patterns: (i) hopping once and then hopping back to the original configuration and (ii) fully exchanging the two fermions with each other and. Finally, we apply the partial creation operator $R$ again and measure the resulting fermion density. When the fermions are returned to their original positions, the state remains unchanged, and the second $R$ operator completes the fermion-pair creation, yielding $\ket{\twofermion}$ as expected. However, when the particles are exchanged, we observe the absence of fermions at the end, consistent with the system returning to the vacuum state, $\ket{\nofermion}$. This represents direct demonstration of the superposition state transforming into $(\ket{\nofermion}{-}\ket{\twofermion})/\sqrt{2}$ due to the particle exchange, which leads to destructive interference with the final creation operation.

To quantify this process, we evaluate the difference in average density on the target sites between the outcomes of the two protocols, corresponding to the contrast resulting from the fermionic Ramsey sequence, under varying levels of error detection (Fig.~\ref{fig:fig4}g). We find that the accuracy of the exchange statistics improves as the quality of fermion encoding is enhanced through decoding postselection.

\naturepar{Simulating the Fermi-Hubbard model}
The Fermi-Hubbard model is one of the most important strongly interacting fermionic systems. In its simplest form it consists of spin-1/2 particles with contact interactions between different fermionic spin states,
\begin{equation}
    H_{\rm FH} = \sum_{\sigma\in{\uparrow,\downarrow}}\sum_{\braket{i,j}}(a^\dagger_{\sigma,i}a_{\sigma,j}+a^\dagger_{\sigma,j}a_{\sigma,i}) + U\,n_{\uparrow}n_{\downarrow},\label{eq:fh}
\end{equation}
where $\sigma$ denotes the spin state, $\braket{i,j}$ is the set of edges encoding the system connectivity, and $U$ is the interaction energy. This model, and its various extensions, provides a description of real-world materials and is believed to underpin some of the most puzzling phenomena in condensed matter, such as high-temperature superconductivity~\cite{lee_doping_2006}.

\begin{figure*}
\centering
\includegraphics[width=180mm]{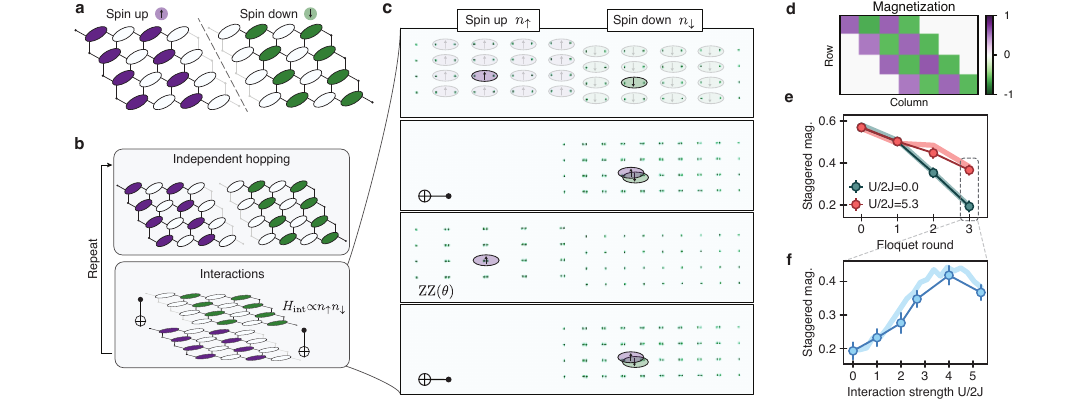}
\caption{\textbf{Two-component strongly interacting fermions.} \textbf{a,} Two independent species of fermions, labeled ``spin-up" and ``spin-down", are realized by dividing the system into two halves. The XX and YY links across the boundary are turned off to facilitate independent hopping dynamics; this system realizes a 4x4 lattice with two fermion species. \textbf{b,} Floquet evolution consists of independent hopping within the two halves followed by density-density interactions between the two spin states. \textbf{c,} The interactions are realized through a parallel entangling operation connecting the two halves of the system. Two parallel CNOT gates propagate a ZZ($\theta$) operator, applied to the spin-up half, into the interaction term.
\textbf{d,} The initial state is a checkerboard pattern, staggered between the two fermion spin states with mean magnetization denoted by the colorbar. \textbf{e,} Staggered magnetization dynamics during a quench with and without the interactions. \textbf{f,} Final staggered magnetization as a function of interaction strength. Increasing interactions  suppress the spin mixing dynamics. The solid lines are the result of noisy numerical simulations of the circuit (Methods). Error bars represent 68\% confidence intervals.}
\label{fig:fig5}
\end{figure*}

To simulate this model, we split our array into two halves, representing spin up on the left side and spin down on the right side, and engineer interactions by realizing parallel gates between the two halves (Fig.~\ref{fig:fig5}a). After the initial encoding step, we omit the hopping terms across the boundary during evolution, enabling independent dynamics of the two spin states. Furthermore, we extend our Floquet cycle by adding a fermion interaction term (Fig.~\ref{fig:fig5}b). Effectively, this realizes a Fermi-Hubbard model on a 16-site square lattice. 

More specifically, to implement onsite density-density interactions, $H_{\rm int}{\propto}n_{\uparrow}n_{\downarrow}$, we take advantage of long-range atom motion to couple distant parts of the system (Fig.~\ref{fig:fig5}c). We apply two parallel CNOT gates across the two halves of the array, interleaved with evolution under the $K^{\rm Z}_{ij}$ interaction on the target half only. This $K^{\rm Z}_{ij}$ term is proportional to ${\rm Z}_{\uparrow} {\rm Z}_{\uparrow}$, and the CNOT gates propagate it to a four-body $({\rm Z}_{\uparrow} {\rm Z}_{\uparrow}){\times}({\rm Z}_{\downarrow} {\rm Z}_{\downarrow})$ term across the two halves, which effectively realizes contact interactions ($4n_{\downarrow}n_{\uparrow}{-}2n_{\downarrow}{-}2n_{\uparrow}$) between the two fermion species.

We initialize an antiferromagnetic checkerboard spin ordering (Fig.~\ref{fig:fig5}d) with local single-qubit gates (Extended Data Fig.~\ref{fig:ED_FermiHubbard}a), and study its dynamics with varying interaction strength. In the absence of interactions ($U{=}0$), we observe that the staggered magnetization of the initial state quickly decreases after a few Floquet cycles, due to the independent fermion hopping (Fig.~\ref{fig:fig5}e). However, by increasing the interactions, we observe suppressed particle transport and significantly slower decay of the staggered magnetization (Fig.~\ref{fig:fig5}e). We measure the staggered magnetization after the final Floquet round for a range of interaction strengths (Fig.~\ref{fig:fig5}f), and find good agreement with numerical simulations of our circuit (Methods), including a non-monotonic behavior as a function of interaction strength.

\naturepar{Outlook}
Our experiments demonstrate key building blocks for digital quantum simulations of complex fermionic and topological systems. Utilizing measurement-based preparation of long-range entanglement, efficient Floquet dynamics, and error detection, they pave the way for explorations of complex and practically relevant quantum systems. These studies can be extended in a number of interesting directions. For instance, properties of the non-Abelian excitations~\cite{kalinowski_non-abelian_2023} can be explored in the Kitaev model, including fusion and braiding properties~\cite{kitaev_anyons_2006}. Moreover, the stability of spin liquid phases can be probed in challenging-to-simulate regimes  by adding different types of interactions, which can be implemented with additional atom moves. Such simulations have the potential to provide new insights into the understanding of so-called Kitaev materials~\cite{Takagi.2019}. Similar methods can be extended to study a broad range of physics and chemistry models, including those relevant to materials and molecules~\cite{maskara_programmable_2023,clinton_towards_2024}, lattice gauge theories~\cite{cochran_visualizing_2024}, and quantum gravity~\cite{maldacena2023simplequantumdescribesblack,sahay_gravity_2024}. Other fermionic encodings can also be explored, either using qubits~\cite{nigmatullin_experimental_2024} or hybrid analog-digital architectures with fermionic atoms~\cite{hartke_fermionpairs_2022,cuadra_fermion_2023}.

While the methods used in this work can readily be scaled up to thousands of qubits~\cite{manetsch_tweezer_2024,jwpan_group_large_rrays_2024}, the achievable circuit depths in digital quantum simulations will be limited due to errors. Our experiments demonstrate that error detection approaches can be used to suppress errors in near-term simulations. Eventually, error correction will need to be integrated to further extend achievable circuit depths. The topological state-preparation circuit can be realized with encoded logical qubits using transversal operations~\cite{zhou_algorithmic_2024}, while implementation of specialized error-correcting codes could further reduce resource overheads~\cite{xu_fast_2024}. The techniques developed and demonstrated in this work provide the foundation for co-designing such simulation approaches in a hardware-efficient manner.

\deleted{
During the completion of this work, we became aware of related work studying Kitaev's honeycomb model on a superconducting quantum computer~\cite{google_kitaev_paper}.}

\FloatBarrier
\bibliographystyle{naturemag_arxiv}
\bibliography{library}

\clearpage
\newpage

\section*{Methods}

\naturepar{Experimental system}
We use the experimental apparatus previously described in Refs.~\cite{bluvstein_quantum_2022,evered_high-fidelity_2023,bluvstein_logical_2024,ebadi_quantum_2021,manovitz_quantum_2024}, with key upgrades that enable efficient digital simulation of Hamiltonian systems. $^{87}$Rb atoms are stochastically loaded from a magneto-optical trap into programmable configurations of 852-nm traps generated with a spatial light modulator (SLM, Hamamatsu X13138-02), and then rearranged with 852-nm moving traps generated by a pair of crossed acousto-optic deflectors (AODs, DTSX-400, AA Opto-Electronic) to realize defect-free arrays~\cite{barredo_atom-by-atom_2016, scholl_quantum_2021, ebadi_quantum_2021}. We image atoms with a 0.65-NA objective (Special Optics) onto a CMOS camera (Hamamatsu ORCA-Quest C15550-20UP). The qubit state is encoded in $m_F\,{=}\,0$ hyperfine clock states of the $^{87}$Rb ground-state manifold (coherence time $T_2{>}1$s~\cite{bluvstein_quantum_2022}), and two-photon Raman excitation is used to drive fast, high-fidelity single-qubit gates ~\cite{bluvstein_quantum_2022,levine_dispersive_2022}. We use both a global Raman path to drive rotations on the entire array and local Raman light generated by 2D AODs and sent through the objective~\cite{bluvstein_logical_2024}. The local single-qubit Raman gates are realized through two different schemes~\cite{bluvstein_logical_2024}, either through local Z rotations or local X rotations. For the feedforward local pulses and the local pulses used to prepare patterns of fermions (in Figs.~\ref{fig:fig4}~and~\ref{fig:fig5}), we use local single-qubit Z gates~\cite{bluvstein_logical_2024}. For the measurement-based state preparation circuit and final local measurement basis rotations, we use local Raman X rotations as will be described in more detail in Ref.~\cite{other_aa_paper}. To realize high-fidelity entangling gates~\cite{evered_high-fidelity_2023,ma_high-fidelity_2023,tsai_gates_2024, cao_multi-qubit_2024, muniz_high-fidelity_2024}, we excite the atoms to the $n\,{=}\,53$ Rydberg state using a two-photon scheme with 420-nm and 1013-nm lasers~\cite{levine_parallel_2019}. We use a closer intermediate state detuning of 3.3\,GHz compared to our previous work~\cite{evered_high-fidelity_2023,bluvstein_logical_2024}. This allows us to address a larger gate region and reduce detuning inhomogeneity from the 1013-nm light shift to improve uniformity of the entangling gates across the array, at the cost of slightly higher scattering~\cite{evered_high-fidelity_2023}. Between entangling gates, we rearrange the atoms dynamically with the AOD traps to achieve any-to-any connectivity~\cite{bluvstein_quantum_2022,bluvstein_logical_2024}.

In our experimental layout (Extended Data Fig.~\ref{fig:ED_ExperimentSequence}a), the hexagonal plaquettes of the honeycomb model are embedded in a rectangular atom array with 4 rows. In these experiments, we use three separate zones: entangling, storage, and readout~\cite{bluvstein_logical_2024}. Atoms are first sorted in the storage zone, then transported into the entangling zone before the start of the experiment. We employ an upgraded sorting algorithm, which, compared to our previous algorithm~\cite{ebadi_quantum_2021}, has an additional final step to fill individual defect sites from a small atom reservoir (more details in Ref.~\cite{other_aa_paper}). For the state preparation circuit, we perform midcircuit measurement~\cite{deist_mid-circuit_2022,singh_mid-circuit_2023,graham_midcircuit_2023,lis_midcircuit_2023,norcia_midcircuit_2023,ma_high-fidelity_2023} of the ancilla qubits, by bringing them to a spatially separated readout zone far away (${\sim}150\mu$m) from the entangling zone and imaging them with a single, local 780-nm imaging beam~\cite{bluvstein_logical_2024}. The fidelities of individual components such as single-qubit gates and two-qubit CZ gates in this work are roughly similar to our other works~\cite{evered_high-fidelity_2023,bluvstein_logical_2024}, except for the local imaging fidelity, which was lower during data taking due to degrading trap laser power ($\sim$96-97\% compared to 99.8\% in Ref.~\cite{bluvstein_logical_2024}).

\naturepar{Tunable entangling phase gates}
A key upgrade to our experiment, enabling efficient digital evolution, is the use of tunable entangling controlled-phase gates (denoted CPHASE or CP) characterized by the angle $\theta$ (normalized such that $\theta{=}1$ is the CZ gate). We implement each CP gate using a single Rydberg laser pulse with constant intensity and 
a phase profile given by the cosine function, $A\cos(\omega t+\varphi)$, and a constant two-photon detuning $\delta$~\cite{evered_high-fidelity_2023}. These parameters are numerically calculated for each $\theta$ using optimal control methods~\cite{evered_high-fidelity_2023,jandura_time-optimal_2022,pagano_gates_2022}. 
For Floquet evolution, we combine two CP gates with global single-qubit $\rm{X}$ operations to realize entangling gates of the form,
\begin{equation}
    {\rm ZZ}(\theta)=e^{i\theta\frac{\pi}{4} {\rm Z}\otimes{\rm Z}}={\rm CP}[\theta/2]\,({\rm X}{\otimes} {\rm X})\,{\rm CP}[\theta/2]\,({\rm X}{\otimes} {\rm X}),
\end{equation}
which are related to CP gates through single-particle terms. 
This approach is not only a robust way to remove the single-qubit terms but also ensures that atoms in the entangling zone that are not undergoing gates do not pick up a spurious phase. 

To calibrate the entangling gates, we adapt an approach from Ref.~\cite{mi_timecrystal_2022} which allows for measuring the entangling phase (Extended Data Fig.~\ref{fig:ED_EntanglingGates}c). We first initialize a pair of atoms, one in $\ket{+}_y {=}(\ket{0}{+}i\ket{1})/\sqrt{2}$ and the other in $\ket{0}$. Then we apply a series of gates to the atom pair, which causes the atom in $\ket{+}_y$ to acquire a phase according to the magnitude of the entangling phase. Finally, we apply a single-qubit rotation in the appropriate axis to the atom initially prepared in $\ket{+}_y$ to bring it to $\ket{0}$ for a perfect gate. After expelling atoms in $\ket{1}$ with resonant pushout light, both atoms should be present only if the gates are perfect. This benchmarking sequence is sensitive to both the entangling phase $\theta$ and loss from the gate operation.

To calibrate the gates, we run this circuit with a fixed number of entangling gates, and scan gate parameters to optimize the gate performance on the experiment, in a similar approach to our CZ gate calibration described previously in Ref.~\cite{evered_high-fidelity_2023}. We measure the return probability after 20 CPHASE gates, for different values of the entangling phase (Extended Data Fig.~\ref{fig:ED_EntanglingGates}d). The return probability is higher for smaller entangling phases, owing to the shorter gate duration and reduced average Rydberg population.

\naturepar{Automated calibration}
Due to the large range of gate angles used in this work, we employ automated calibration routines which enable convenient calibration, either using an automated version of the parabola scan method previously used~\cite{evered_high-fidelity_2023} or a Nelder-Mead optimization algorithm~\cite{nelder_simplex_1965} adapted for noisy data~\cite{huang_robust_2018}. Additionally, we perform automated calibration of the Rydberg beams, due to the importance of beam homogeneity for realizing uniform gate angles across the array. We use a flat tophat intensity profile generated using an SLM to maximize homogeneity across the array~\cite{ebadi_quantum_2021}. To address local deviations in the beam intensity, we use pre-calculated ``peak correction" phase masks which, upon addition to the base hologram, realize localized intensity peaks (Extended Data Fig.~\ref{fig:ED_EntanglingGates}f). We use automated noisy Nelder-Mead optimization to sequentially adjust both Zernike and peak corrections on the SLM. An example calibration is shown in Extended Data Fig.~\ref{fig:ED_EntanglingGates}g, where the peak-to-peak variation in beam intensity between rows in the array (as measured by the light shift on the hyperfine qubit) decreases throughout calibration. This active optimization procedure significantly improves the homogeneity of the intensity profile across the rows (Extended Data Fig.~\ref{fig:ED_EntanglingGates}h), and can be used in combination with passive approaches for removing aberrations~\cite{zupancic_ultra-precise_2016,ebadi_quantum_2021,chew_ultra-precise_2024}.

\naturepar{Measurement-based preparation of topological states} 
We begin our experiments by efficiently preparing a long-range entangled, topological state on a cylinder, which forms the basis for all subsequent explorations. The high-level description of the procedure is presented in Extended Data Fig.~\ref{fig:ED_AncillaMeasurement}a-b. First, we prepare a ZXXZ surface code state~\cite{kitaev_fault-tolerant_2003,bonilla_ataides_xzzx_2021} on one of the data qubit sublattices (black sites), using entanglement operations with ancilla qubits to project the data-qubit state into an eigenstate of the weight-4 ZXXZ operators. Concretely, we put the ancilla qubits in the $\ket{0}_a{+}\ket{1}_a$ state and perform a sequence of CZ gates, with data qubits in the appropriate bases, which realizes the $\ket{0}_a{+}{\rm Z}_1{\rm X}_2{\rm X}_3{\rm Z}_4\ket{1}_a$ state, where the numbers label the data qubits along an example ZXXZ operator (Extended Data Fig.~\ref{fig:ED_AncillaMeasurement}a). The subsequent measurement of the ancilla qubits in the X basis fixes the parity of the ZXXZ operators on the data qubits. The measurement outcomes and, thus, the projected operator parities are random (up to certain constraints on their products). A feedforward step, acting on the same sublattice, ensures that all parities end up being +1, which corresponds to the subspace with the ground state of the Kitaev model~\cite{lieb_flux_1994}. For the fermion dynamics, this ensures that no magnetic fluxes are present.

The order of entangling bases we use is Z, X, X, Z and since all qubits are initially in the $\ket{0}$ state, we can omit the first Z measurement; the resulting circuit for this part is depth-3. These weight-4 ZXXZ operators are grown to weight-6 ZYXZXY operators by performing parallel controlled-Y (CY) operations between the entangled sublattice (black) and the one remaining in the $\ket{0}$ state (green). The resulting weight-6 operators are equivalent to the plaquettes $W_p$=ZYXZYX up to the ZZ-link operators (Extended Data Fig.~\ref{fig:ED_AncillaMeasurement}a). However, the structure of the circuit ensures that the ZZ-link terms are +1 and, thus, can be freely multiplied into other operators. As a result, this depth-4 circuit efficiently prepares the fermionic vacuum sate, or equivalently the A$_{\rm Z}^{\rm I}$ ground state of the Kitaev honeycomb model~\cite{kitaev_anyons_2006}. Such measurement-based methods can be used for preparing a wide range of topological states in finite depth~\cite{lu_measurement_2022,tantivasadakarn_hierarchy_2023,sahay_finite-depth_2024}.

The experimental implementation of this sequence is shown in Extended Data Fig.~\ref{fig:ED_AncillaMeasurement}c-e. Initially, the ancilla qubits are located in moveable AOD traps and the data qubits are in stationary SLM traps. The ancilla qubits are reconfigured and then entangled with data qubits in the correct basis. The first entanglement step includes the periodic boundary direction and has an additional component where the top row of ancilla qubits is entangled with the bottom row of data qubits. We perform this step first so that all other qubits can be in state $\ket{0}$, avoiding additional errors due to Rydberg excitations.

To realize midcircuit measurement, the ancilla qubits are transported to the readout zone and locally imaged~\cite{bluvstein_logical_2024}. The measurement outcomes are then used to perform real-time decoding and feedforward. The feedforward correction is applied before the parallel CY operations (Extended Data Fig.~\ref{fig:ED_AncillaMeasurement}e).

For the feedforward corrections, we use a field-programmable gate array, FPGA (Xilinx ZCU111), to gate on and off 32 local \deleted{Raman} \new{single-qubit} Z gates applied across one sublattice of the array (Extended Data Fig.~\ref{fig:ED_AncillaMeasurement}d). An example of mid-circuit decoding and feedforward is shown in ED Fig.~\ref{fig:ED_FeedforwardMethods}a. The decoding algorithm uses single-site Z gates, which flip the two vertically adjacent plaquettes, to pair the -1 results in each column by pushing the -1 values until another one is encountered. The initial site for each column and the direction of the pushing procedure is randomized to avoid biasing any given row of plaquettes. The decoder can additionally be modified to prepare an initial state with a deterministic pattern of $\pm$1 plaquette values, as long as the configuration does not violate parity constraints. In Fig.~\ref{fig:fig5}, we use this flexibility to initialize different states with fewer local pulses than naively necessary: if a single-qubit gate pattern used to initialize the fermion sites flips an even number of plaquettes per column, we can pre-compensate for those flips in the decoding step (Extended Data Fig.~\ref{fig:ED_FermiHubbard}a).

In Extended Data Fig.~\ref{fig:ED_FeedforwardMethods}d-e, we numerically explore various constant-depth circuits realizing the long-range entangled state of interest. The method we employ in this work significantly outperforms other approaches, including the direct measurement of the hexagonal plaquettes.

\naturepar{Decoding error detection}
The product of ancilla measurement outcomes in every column must be even since the plaquettes in any column multiply to strings enclosing the cylinder that are composed of Z operators only (Extended Data Fig.~\ref{fig:ED_FeedforwardMethods}b), which are fixed to be +1 due to our initial product state. This constraint can be used for error detection. In particular, whenever there are an odd number of -1 ancilla measurement results in a given column, we know that an error must have occurred during the state preparation circuit. Utilizing this decoding postselection method, we observe that all lengths of loops improve in value, as shown in Extended Data Fig.~\ref{fig:ED_FeedforwardMethods}c.

Throughout this work, we find that error detection is not necessary for achieving the main results but consistently improves data quality. We define a \textit{decoding threshold} for the number of columns that can have odd ancilla parity. For example, a decoding threshold of 0 means that all 8 columns have the correct even ancilla parity. In Fig.~\ref{fig:fig3}a we use a decoding threshold of 1, and in Fig.~\ref{fig:fig3}c, a decoding threshold of 2. In Fig.~\ref{fig:fig4}c, we do not use decoding postselection, and in Fig.~\ref{fig:fig4}e, we used a decoding threshold of 4. For Fig.~\ref{fig:fig4}f, we use a decoding threshold of 1, except for the full exchange final state, for which we use a decoding threshold of 0. Finally, for Fig.~\ref{fig:fig5}e-f, we use a decoding threshold of 1.

\naturepar{Postselection based on atom loss}
We utilize a state-selective qubit readout\new{, described in our work in Ref.~\cite{other_aa_paper},} which distinguishes between the $\{\ket{0},\ket{1}\}$ states and atom loss (with the exception of mid-circuit ancilla measurement and the data in Fig.~\ref{fig:fig2} and Extended Data Fig.~\ref{fig:ED_FeedforwardMethods}c). \deleted{This method is based on converting the atomic state to its spatial position~\cite{wu_stern-gerlach_2019} and is covered in detail in our upcoming work~\cite{other_aa_paper}.} \new{In this method, a state-dependent circularly-polarized 1D optical lattice is used to pin one of the two qubit states~\cite{robens_atomic_2016,mandel_coherent_transport,wu_stern-gerlach_2019}, and an AOD tweezer moves the other qubit state a few microns away \cite{other_aa_paper}, converting the atomic state to position which is then imaged via polarization gradient cooling (PGC).} The information about lost atoms does not allow us to correct the state but we can use it for error detection and postselection~\cite{scholl_erasure_2023,ma_high-fidelity_2023}. In particular, for each observable, we only use the experimental shots where all qubits constituting that observable are present. Moreover, we can employ a sliding-scale postselection scheme where we additionally postselect on qubits being present within a given distance on the lattice (Extended Data Fig.~\ref{fig:ED_ExperimentSequence}b), which can mitigate the effects of error spreading throughout the circuit. To quantify this, we introduce a \emph{loss radius}, which describes the distance on the honeycomb lattice within which the atoms must be present.

We use varying amounts of loss postselection for different results in this work. Fig.~\ref{fig:fig2} has no loss postselection because we do not use state-selective readout for the data in that figure. In Fig.~\ref{fig:fig3}a, we use a loss radius of 0, and in Fig.~\ref{fig:fig3}c, a loss radius of 2 for all the string observables. 
In Fig.~\ref{fig:fig4}c,e, we use a loss radius of 0, and in Fig.~\ref{fig:fig4}f, a loss radius of 2 for all the exchange color plots. Finally for Fig.~\ref{fig:fig5}e-f, we use a loss radius of 2.

Here we briefly summarize the acceptance fraction for data throughout the paper, when using the loss and decoding postselection methods. We emphasize that this postselection is not critical to the main results of the paper, but rather we use this tool to improve the quality of results and elucidate the phenomena. In Fig.~\ref{fig:fig2}\deleted{c}\new{d}, the acceptance fractions are directly plotted on the x-axis (there is no additional overhead of loss detection because we do not use state-selective readout for this data). Similarly, for Fig.~\ref{fig:fig2}\deleted{d}\new{c}, the acceptance fractions are plotted in Extended Data Fig.~\ref{fig:ED_FeedforwardMethods}c for the different loops. For the string observables plotted in Fig.~\ref{fig:fig3}c, the acceptance fraction lies in the range of $3\%$ to $25\%$. In Fig.~\ref{fig:fig4}c, the acceptance fraction is in the range from $65\%$ to $86\%$ (with higher acceptance for the shorter depths due to less loss from gates), and for the data in Fig.~\ref{fig:fig4}c is $61\%$. In Fig.~\ref{fig:fig4}f, for the fermion exchange color plots, it is in the range of $1\%$ to $7\%$ and in Fig.~\ref{fig:fig4}g, the acceptance fraction is shown in the x-axis. Finally, for Fig.~\ref{fig:fig5}e, it ranges from $2\%$ to $5\%$ depending on the Floquet round, with all the points in Fig.~\ref{fig:fig5}f being $2\%$.

\naturepar{Floquet evolution circuits}
After the measurement-based state preparation, we keep one data qubit sublattice in moveable AOD traps (denoted by black circles in Extended Data Fig.~\ref{fig:ED_AncillaMeasurement}), and at each time step, transport them next to the atoms in the even data qubit sublattice to perform the tunable entangling gates. Global single-qubit rotations between entangling gates are used to change the basis between X, Y, and Z. We perform dynamical decoupling throughout, including to cancel single-qubit dephasing from moves~\cite{bluvstein_quantum_2022,bluvstein_logical_2024}. During the Floquet evolution, we implement periodic cylindrical boundary conditions. In particular, due to our array geometry, only the XX links couple the top row to the bottom row. For these links, we apply it in two steps, first moving the data qubits in AOD traps up one lattice site, and then moving the top row to perform gates with the bottom row (Fig.~\ref{fig:fig1}c). During this second step, the three other rows are moved out of the Rydberg beam to avoid extra gate errors. Conveniently, the structure for all the Floquet circuits in this work is the same (other than in Fig.~\ref{fig:fig5}), and so we only need to change the CPHASE gates between the different Floquet circuits. This enables the implementation of a wide variety of fermionic evolution with minimal experimental changes.

\new{Since the Floquet circuit commutes with closed loop operators, the topological order is preserved as we evolve into phase B, up to a small decay from gate errors (Fig.~\ref{fig:fig3}b). In principle, the final state could be more fully characterized by measuring all closed loop operators, including the additional larger loops measured in the initial state (Fig.~\ref{fig:fig2}b).}

For the low-energy states of the Kitaev Hamiltonian in Fig.~\ref{fig:fig3}, we optimize the state preparation circuits to maximize the overlap between bulk Majorana correlations (Extended Data Fig.~\ref{fig:ED_MajoranaChern}a-b) of the prepared state and that of the Floquet ground state with Floquet time $\tau{=}1/4J$. We confirm \deleted{independently} that the resulting state has \new{the correct Chern number and a} similar energy as a state optimized purely based on energetic considerations. \new{We note that the circuit depth of 6 is sufficient for this system size, but will necessarily increase for larger system sizes due to the gap closing~\cite{kalinowski_non-abelian_2023}}. For the non-Abelian circuit, we used the CPHASE gates as shown in Fig.~\ref{fig:fig3}b, and for the Abelian II circuit, the circuit requires only three CPHASE gates: CP[-0.0625], CP[-0.0625], and CP[-0.3125]. The single-qubit sequence and atom motion is the same between all circuits in Fig.~\ref{fig:fig3}, with the exception of the final measurement basis rotations. The value of the plaquette parity at depth 0 is lower in Fig.~\ref{fig:fig3}b than in Fig.~\ref{fig:fig2} due to these additional errors.

For Fig.~\ref{fig:fig4}c, we use CP[-0.0625] gates to realize $J\tau{=}0.125$, and for the case $J_{\rm Z}/J{=}8$, we use CP[0.5] gates for the $J_{\rm Z}$ term. For Fig.~\ref{fig:fig4}d and Fig.~\ref{fig:fig5}, we use CP[-0.0938] gates for the $J_{\rm X}$ and $J_{\rm Y}$ terms and CP[0.5] gates for the $J_{\rm Z}$ term in order to slightly increase the amount of hopping dynamics. We perform the measurements in Fig.~\ref{fig:fig4}e after depth-11 (note that the final ZZ-term for depth-12 commutes with the Z-basis measurement so we omit the final circuit layer).

\naturepar{Encoding fermions in qubits}
The Hilbert space of $N$ qubits and $N$ fermions has the same dimension but due to non-local properties of fermions (anticommutation relations), mapping between them can be complicated.
A direct translation on the operator level is given by the Jordan-Wigner (JW) transformation~\cite{jordan_ber_1928}, where for a particular site ordering, $[1,...,N]$, we can identify complex-fermion creation and annihilation operators
\begin{align*}
    a_j &= \frac{1}{2}({\rm X}_j+i{\rm Y}_j) \prod_{k=1}^{j-1} {\rm Z}_k, \\
    a^\dagger_j &= \frac{1}{2}({\rm X}_j-i{\rm Y}_j) \prod_{k=1}^{j-1} {\rm Z}_k, &
\end{align*}
and the corresponding Majorana operators,
\begin{align*}
    c_j &= a_j^\dagger{+}a_j = {\rm X}_j \prod_{k=1}^{j-1} {\rm Z}_k, &
    \bar{c}_j &= i(a_j^\dagger{-}a_j) =  {\rm Y}_j \prod_{k=1}^{j-1} Z_k,
\end{align*}
which can be directly checked to satisfy the canonical anticommutation relations, $\{a_i,a_j^\dagger\}{=}\delta_{ij}$ and $\{c_i,c_j\}{=}2\delta_{ij}$. Beyond one dimension, this approach leads to macroscopic operator weight on the qubit side, even for simple local fermion operations such as nearest-neighbor hopping, 
\begin{equation}
    c_i c_j={\rm X}_i{\rm X}_j \prod_{k=i}^{j-1} Z_k,\label{seq:jwhop}
\end{equation}
 for $i{<}j$ (Extended Data Fig.~\ref{fig:ED_FermionEncoding}a).

The idea of local fermion-to-qubit encodings~\cite{verstraete_mapping_2005} relies on introducing a long-range entangled ``background state" $\ket{\nofermion}$ that is stabilized by the non-local operator strings, effectively canceling them out. For example, if $\prod_k Z_k$ in Eq.~\eqref{seq:jwhop} is a stabilizer of $\ket{\nofermion}$, i.e. $(\prod _k Z_k )\ket{\nofermion}{=}\ket{\nofermion}$, then the hopping term effectively becomes a simple weight-2 operator, $c_ic_j{\sim}{\rm X}_i {\rm X}_j$. 
However, introducing these additional constraints on the state reduces the available Hilbert space for fermion degrees of freedom; thus, the system needs to be expanded by introducing additional data qubits. In the honeycomb encoding studied here, this manifests as the long-range entangled ZXXZ state on one sublattice being coupled to matter degrees of freedom through the parallel controlled-Y operation (Fig.~\ref{fig:fig2}a). In this setting, we have twice as many qubit degrees of freedom compared to fermionic ones, which grants enough space to  enforce the stabilizer constraints.

We now show that the two-qubit interactions introduced in Eq.~\eqref{eq:spinint} correspond to hopping terms of Majorana fermions, $c_j$, localized at each site $j$. We begin by constructing JW operators that are similar to those in Eq.~\eqref{seq:jwhop} but modified to better suit our honeycomb lattice. 
We choose a site ordering starting in the top-left corner of the lattice and creating a continuous path $\mathcal{L}$ through the entire system; for example, like the one in Extended Data Fig.~\ref{fig:ED_FermionEncoding}a. We use a JW operator~\cite{kells_description_2009},
\begin{equation*}
    c_j = {\rm Z}_1\prod_{l\in \mathcal{L}_{j}} \sigma^{(l)}_{l(1)} \sigma^{(l)}_{l(2)},
\end{equation*}
where $\mathcal{L}_{j}$ is the sequence of links along path $\mathcal{L}$ ending at site $j$, $l(1/2)$ denotes the vertices of link $l$, and $\sigma^{(l)}{\in}\{{\rm X},{\rm Y},{\rm Z}\}$ is the Pauli operator along $l$. Since the consecutive link operators always anticommute, the $c_j$ operators defined this way satisfy the correct anticommutation relations and, as products of Paulis, are Hermitian and square to the identity operator. Now, if the hopping terms is between two consecutive sites on path $\mathcal{L}$, then we trivially recover $\sigma^{(l)}_{l(1)} \sigma^{(l)}_{l(2)}$ as the rest of the string squares to identity. Finally, if the hopping term is between two nearest-neighbor sites that are not adjacent on  $\mathcal{L}$, we end up with 
\begin{align*}
    c_i c_j = \prod_{l\in \mathcal{L}_j\setminus \mathcal{L}_i} \sigma^{(l)}_{l(1)}\sigma^{(l)}_{l(2)},
\end{align*}
where $\mathcal{L}_j{\setminus}\mathcal{L}_i$ is the ordered set of links between sites $i$ and $j$ along $\mathcal{L}$. However, by applying the link $(i,j)$, that path can be completed to a closed loop (on a cylinder it also needs to be multiplied by a conserved non-trivial loop around the cylinder) which, in turn, is exactly the product of all enclosed plaquette operators (Extended Data Fig.~\ref{fig:ED_FermionEncoding}c). If the enclosed plaquettes are +1, the hopping term is effectively $\sigma^{(l)}_{i} \sigma^{(l)}_{j}$, where $l{=}(i,j)$. Thus, if all plaquettes are projected to be +1, such nearest-neighbor terms become the link operators and more general Majorana correlations are mapped to a Pauli string constructed from products of link operators. The plaquettes with -1 values act as $\mathbb{Z}_2$ magnetic field fluxes, since they result in a $\pi$-phase for fermions hopping around them.

Complex fermions can always be formed by arbitrary pairing of Majoranas and here we choose to combine them along the ZZ links. Extended Data Fig.~\ref{fig:ED_FermionEncoding}d shows how the hopping operator for such complex fermions can be realized through a linear combination of length-2 and length-4 Pauli strings, which symmetrically couple the different Majorana constituents. The operators of this form can be realized through Floquet engineering. 

The two-qubit Pauli operators corresponding to the nearest-neighbor Majorana hopping, as derived here, constitute the exact interactions $K^{\rm X}_{ij}$, $K^{\rm Y}_{ij}$, and $K^{\rm Z}_{ij}$, implemented in this work. Moreover, the Hamiltonian given by nearest-neighbor Majorana hopping terms results in the original Kitaev honeycomb model~\cite{kitaev_anyons_2006}. In Extended Data Fig.~\ref{fig:ED_FermionEncoding}e, we summarize all operators used in this work and explicitly write them out in both the qubit and fermion languages.

\naturepar{Free-fermion states and Hamiltonians}
The Kitaev honeycomb model has an exact solution in terms of free fermions~\cite{kitaev_anyons_2006}, which enables efficient numerical simulation and benchmarking of most circuits in this work. Here we briefly summarize the main properties of such free-fermion states and Hamiltonians, focusing on Majorana operators, $c_i$, satisfying the canonical anticommutation relations $\{c_i,c_j\}{=}2\delta_{ij}$.

Free-fermion states are captured by the two-point correlation matrix $\bm{\Gamma}$,
\begin{equation}
    \Gamma_{ij} = \frac{i}{2}\braket{[c_i,c_j]},\label{seq:majcor}
\end{equation}
with all higher-order terms decomposing into products of two-point functions via Wick's formula~\cite{wick_evaluation_1950}. 
A general quadratic Majorana Hamiltonian,
\begin{equation}
    H = \frac{i}{4} \sum_{} c_i A_{ij} c_j,\label{seq:majham}
\end{equation}
is defined through a real, skew-symmetric matrix $\bm{A}^\top{=}{-}\bm{A}$, where matrix elements $A_{ij}$ encode Majorana hopping between sites $i$ and $j$. The correlations of the ground state are related to those of the Hamiltonian~\cite{kitaev_anyons_2006}, and a unitary evolution of an arbitrary free-fermion state $\bm{\Gamma}$ is given by,
\begin{equation}
    \bm{\Gamma}(t) = \bm{U}^\dagger(t)\bm{\Gamma}(0) \bm{U}(t), \label{seq:maju}
\end{equation}
where $\bm{U}(t){=}\exp(-\bm{A} t)$ is a matrix describing the time-evolution of the two-point correlation matrix~\cite{kraus_generalized_2010}.

We focus on a translationally invariant system, which can be described by the unit-cell position $\bm{R}$, and a label for the site within that unit cell, $\lambda$. For the honeycomb lattice, the unit cell has two sites, $\lambda{=}\{e,o\}$, corresponding to the even and odd sublattices, respectively, and Eq.~\eqref{seq:majham} may be re-written as,
\begin{equation}
    H = \frac{i}{4} \sum_{\bm{R},\bm{r}} \begin{bmatrix}
        c^{(e)}_{\bm{R}} & c^{(o)}_{\bm{R}}
    \end{bmatrix}  \begin{bmatrix}
        A^{(e,e)}_{\bm{r}} &A^{(e,o)}_{\bm{r}} \\
        A^{(o,e)}_{\bm{r}} & A^{(o,o)}_{\bm{r}}
    \end{bmatrix} \begin{bmatrix}
        c^{(e)}_{\bm{R}+\bm{r}} \\ c^{(o)}_{\bm{R}+\bm{r}}
    \end{bmatrix},
\end{equation}
where $\bm{r}$ is the relative position between the two relevant unit cells. In terms of momentum modes, $c_{\bm{k}}{\propto}\sum_{\bm{x}} e^{-i\bm{k}\cdot\bm{x}}c_{\bm{x}}$, where ${\bm{x}}$ is the position operator, the Hamiltonian takes the form $H{=}\sum_{\bm{k}}H_{\bm{k}}$,
\begin{equation}
    H_{\bm{k}} = \frac{i}{4}\begin{bmatrix}
        c^{(e)}_{\bm{-k}} & c^{(o)}_{\bm{-k}}
    \end{bmatrix} \begin{bmatrix}
        \Delta_{\bm{k}} &\xi_{\bm{k}} \\
        -\xi_{\bm{k}}^\ast & -\Delta_{\bm{k}}
    \end{bmatrix} \begin{bmatrix}
        c^{(e)}_{\bm{k}} \\ c^{(o)}_{\bm{k}}
    \end{bmatrix},
\end{equation}
where $\Delta$ and $\xi$ functions are  the Fourier transforms of $A_{\bm{r}}^{(e,e)}$ and $A_{\bm{r}}^{(e,o)}$, respectively. The Hamiltonian $H$ contains two energy bands with dispersions $\pm\varepsilon_k{\propto}\sqrt{\xi_{\bm k}^2{+}\Delta_{\bm k}^2}$. In the original Kitaev model, the gap closes in phase B~\cite{kitaev_anyons_2006}, $\Delta_{\bm{k}}{=}0$, but our effective Floquet Hamiltonians have a spectral gap due to natural breaking of the time-reversal symmetry~\cite{kalinowski_non-abelian_2023}. Knowing the numerical values of the $\Delta_{\bm{k}}$,$\xi_{\bm{k}}$ functions on a grid of points in the Brillouin zone enables evaluation of various band properties. In particular, evaluation of the Chern number requires only a few points in momentum space~\cite{fukui_chern_2005}.

\naturepar{Chern number}
The gapped non-Abelian phase B of the Kitaev honeycomb model is characterized by a non-zero Chern number~\cite{berry_quantal_1997,chern_characteristic_1946} of the excitation band. An odd Chern number guarantees that a magnetic flux is accompanied by an unpaired Majorana zero-mode with non-Abelian (Ising anyon) statistics~\cite{kitaev_anyons_2006,teo_topological_2010,alicea_new_2012}. In Fig.~\ref{fig:fig3}, we prepare a vortex-free ground state whose free-fermion parent Hamiltonian has a band with an odd Chern number. In principle, the Majorana zero modes could be prepared and probed directly, but the required circuit depths are larger than those used in this work~\cite{kalinowski_non-abelian_2023}.

The Chern number is a topological invariant of the energy band, which characterizes the geometry of the single-particle eigenstates of $H_{\bm{k}}$~\cite{berry_quantal_1997,cooper_topological_2019}. Those eigenstates, satisfying $H_{\bm{k}}\ket{n_{\bm{k}}}{=}E_{n,\bm{k}}\ket{n_{\bm{k}}}$ for energy $E$, are defined up to an overall choice of phase gauge,
\begin{equation}
    \ket{n_{\bm{k}}}\to e^{-i\phi_{\bm{k}}}\ket{n_{\bm{k}}},\label{seq:nphi}
\end{equation}
where $\phi_{\bm{k}}$ is the gauge parameter. To probe the local geometry of these eigenstates, as the momentum $\bm{k}$ is varied, we define the Berry potential (connection),
\begin{equation}
    A_{n,\bm{k}} = \braket{n_{\bm{k}}|\nabla_{\bm{k}}|n_{\bm{k}}},
\end{equation}
where $\nabla_{\bm{k}}{=}(\partial_{k_x},\partial_{k_y})$ is the gradient in momentum space. The Berry potential is not gauge-invariant, since it transforms as $A_{n,\bm{k}}{\to}A_{n,\bm{k}}{-}i\nabla_{\bm{k}}\phi_{\bm{k}}$ under Eq.~\eqref{seq:nphi}, but the Berry curvature~\cite{berry_quantal_1997},
\begin{equation}
    {\rm F}_{n,\bm{k}} = \nabla_{\bm{k}} \times A_{n,\bm{k}},\label{seq:Fmat}
\end{equation}
is invariant under Eq.~\eqref{seq:nphi} because $\nabla{\times}(\nabla \phi){=}0$.

The Chern number of the $n$th band is the integral of the Berry curvature over the 1st Brillouin zone,
\begin{equation}
    {\rm C}_n = \int_{\rm 1st\, B.Z.}d\bm{k}\, {\rm F}_{n,\bm{k}}\label{seq:chern}
\end{equation}
and is guaranteed to be an integer~\cite{sachdev_quantum_2023}. Throughout this work we focus on the lowest energy band and omit the subscript $n$.
For a given Bloch Hamiltonian in momentum space, $H_{\bm{k}}$, the Chern number can be evaluated either by direct numerical integration of Eq.~\eqref{seq:chern} or with a specialized numerical algorithm~\cite{fukui_chern_2005}. Therefore, the task at hand is reduced to learning the momentum-space parent Hamiltonians of the states prepared in our experiments~\cite{qi_determining_2019,huang_learning_2023,olsacher_hamiltonian_2025}.

The free-fermion states are defined by their two-point correlation functions and, similarly, the free-fermion ground states are related to the single-particle eigenstates of the parent Hamiltonian~\cite{kitaev_anyons_2006}. We measure the open Pauli strings corresponding to the two-point Majorana correlations (Fig.~\ref{fig:fig3}c), and average their values over the bulk region (Extended Data Fig.~\ref{fig:ED_MajoranaChern}a). We include all strings that span the bulk of the system, as depicted in Extended Data Fig.~\ref{fig:ED_MajoranaChern}b, and effectively recover all the $A_{\bm{r}}$ matrix elements of the parent Hamiltonian (up to the norm) with $r_x,r_y{\in}[-1,0,1]$ due to the finite-size restrictions (Extended Data Fig.~\ref{fig:ED_MajoranaChern}c). We then Fourier-transform these correlations onto a regular 5x5 grid in momentum space, resulting in an estimate of the parent Hamiltonian $H{=}{\sum}H_{\bm{k}}$ up to an energy scale $\epsilon_{\bm{k}}$.

Finally, we apply the algorithm of Ref.~\cite{fukui_chern_2005} to evaluate the Chern number. We diagonalize the $H_{\bm{k}}$ Hamiltonians at each $\bm{k}$ and calculate the phases of eigenstate overlaps between neighboring momentum points and collect them in a tensor ${\rm U}^\mu_{\bm k}$, where $\mu{\in}\{\hat{x},\hat{y}\}$ denotes the direction in momentum space, which depends only on the eigenstates of $H_{\bm{k}}$ and not $\epsilon_{\bm{k}}$. Then, we calculate the discretized Berry curvature in Eq.~\eqref{seq:Fmat} (Extended Data Fig.~\ref{fig:ED_MajoranaChern}d) and sum its matrix elements to obtain the Chern number, which is guaranteed to be an integer~\cite{fukui_chern_2005}. \new{An interesting future effort would be to compare this approach to other methods for extracting the Chern number, such as the real-space formula of Ref.~\cite{kitaev_anyons_2006}.}

The Chern number is obtained from the learned Hamiltonian, and cannot be evaluated on individual snapshot data. When applying this procedure with the mean values plotted in the string distributions in Fig.~\ref{fig:fig3}c, we obtain ${\rm C}{=}0$ for the Abelian phase and ${\rm C}{=}1$ for phase B. To study the robustness of this result and the effect of postselection, we bootstrap the data~\cite{davison_bootstrap_1997} by evaluating the Chern number on Hamiltonians learned from randomized subsets (with replacement) of the entire dataset. The averaged Chern number approaches 1 as the batch size grows, and is reduced for small batches due to projection noise (Extended Data Fig.~\ref{fig:ED_MajoranaChern}f). For a batch size above 200, which is still a small fraction of our available data, the averaged Chern number is above 0.9 and quickly approaches 1 with increasing batch size. In that intermediate regime, postselecting on loss leads to a small but noticeable increase in the averaged Chern number (Extended Data Fig.~\ref{fig:ED_MajoranaChern}g). These observations provide further evidence that the Chern number of our output distribution is consistent with having the value of 1.

We further study such Chern number evaluated on a noisy ensemble through numerical simulations, with the same approach as that used to calculate the string distribution in Fig.~\ref{fig:fig3}c. In Extended Data Fig.~\ref{fig:ED_MajoranaChern}h, we evaluate the Chern number on string distributions simulated for various initialization and per-layer errors. We find that our parameters are comfortably within the regime of the unit Chern number.

\naturepar{Numerical simulations with errors}
We perform circuit-level noisy numerical simulations for both the initial state-preparation step and the subsequent Floquet dynamics. The state preparation circuit consists of Clifford operations only, and we simulate it using the \texttt{stim} package~\cite{gidney_stim_2021}. The following Floquet circuit has non-Clifford operations but the effective free-fermion dynamics enable efficient simulation by keeping track of the correlation matrix in Eq.~\eqref{seq:majcor} through unitary dynamics, as described in Eq.~\eqref{seq:maju}.

We incorporate the effect of errors to the numerical simulation of free-fermion systems in a stochastic fashion, with an error channel with strength $p_l$ applied after each circuit layer and the result averaged over many noise realizations. 
The coherent errors simply modify the phase of applied gates. 
The lost qubits are kept track of in a separate data structure and all subsequent gates that include those qubits are removed (with the error model still applied). When evaluating the observables, we postselect the data such that all qubits of the target observable are present (as is done in the experiment).
The stochastic single-site Paulis are also kept track of in a dedicated data structure, and they flip the sign of all the following XX, YY, and ZZ link operators that they anticommute with. At the end of the circuit, the final set of Pauli errors is propagated through the observables being evaluated and flips them accordingly. 
The dynamics of interacting fermions are simulated with the approximate fermionic Gaussian state approach~\cite{kraus_generalized_2010}. 

We use a simple ansatz for our noise per gate layer, which consists of single-qubit incoherent Pauli errors and atom loss, and tune the overall noise strength to match select observables and use those fixed values for the majority of simulations. This phenomenological approach is not directly connected to any particular fidelity, as it models (in a naive way) all experimental contributions.
Since the state-preparation circuit cannot be simulated within the free-fermion framework, we initialize the noisy Floquet simulation with a layer of noise whose strength is chosen to match our experimental plaquette data in Fig.~\ref{fig:fig2}b, which gives the single-qubit initialization error of $p_{\rm ini}{=}0.1$. Similarly, the effective noise per gate layer was calibrated by applying isotropic fermion evolution ($\theta{=}1$) and looking at the ZZ-link observables at depth 12, resulting in the fitted single-qubit error per circuit layer $p_l{=}0.01$. These error rates are divided between atom loss and (unbiased) single-qubit Pauli noise, with the loss constituting $\approx$6\% and 40\% of $p_{\rm ini}$ and $p_l$, respectively. The simulations of circuits with interacting fermions assume perfect gate operations and include initialization errors calibrated to match the initial value in Fig.~\ref{fig:fig5}e. In general, for the fermion simulations (Figs.~\ref{fig:fig4} and \ref{fig:fig5}), we found that the dominant effect of noise was an overall rescaling of quantities rather than a large change in qualitative trends.

\naturepar{Emergent particle number conservation}
In the quench experiment summarized in Fig.~\ref{fig:fig4}a-c, increasing $J_{\rm Z}/J$ leads to emergent particle number conservation at certain time intervals (depths 6 and 12). Here we describe this process in more detail and provide basic derivations of the effective Floquet Hamiltonians. 
The Floquet unitary for a single cycle of the quench experiment is,
\begin{equation}
    U_F = e^{i\sum_{\braket{i,j}_{\rm Z}}K^{\rm Z}_{ij}}e^{i\sum_{\braket{i,j}_{\rm Y}}K^{\rm Y}_{ij}}e^{i\sum_{\braket{i,j}_{\rm X}}K^{\rm X}_{ij}},\label{seq:Uf}
\end{equation}
where the $K_{ij}^{\rm X/Y/Z}$ interactions are defined in Eq.~\eqref{eq:spinint} and $\braket{i,j}_{\rm X/Y/Z}$ are the XX,YY, and ZZ links, respectively. The $K^{\rm Z}_{ij}$ term is, in terms of complex-fermion operators, 
\begin{equation}
    \sum_{\braket{i,j}_{\rm Z}}K^{\rm Z}_{ij} \sim J_{\rm Z}\sum_i n_i = J_{\rm Z}\,N_{\rm tot},
\end{equation}
where $N_{\rm tot}$ is the total particle number operator. Thus, for large $J_{\rm Z}$ values there is a strong term in the Hamiltonian proportional to the total particle number. Since the initial state is an eigenstate of $N_{\rm tot}$, the subsequent evolution is  projected into the subspace of $N_{\rm tot}$ with the same eigenvalue, which can be understood as orthogonal wavefunction components averaging out due to fast oscillations at the timescale of $1/J_{\rm Z}$, effectively preserving the particle number and realizing complex-fermion dynamics. In Floquet evolution, we need at least two applications of $U_F$ for the hopping along XX and YY to be affected by the $N_{\rm tot}$ operator and, thus, the shortest particle-conserving Floquet circuit is depth-6, as seen in Fig.~\ref{fig:fig4}c.

Such intuition can be further substantiated through analytical arguments on the operator level. As we show below, large-angle ${\rm ZZ}(\theta)$ rotations effectively grow the nearest-neighbor link operators to length-4 strings in a way that can realize complex-fermion hopping (Extended Data Fig.~\ref{fig:ED_FermionEncoding}d-e). The effect of varying $J_{\rm Z}$ can be understood by looking at a composite two-site unitary,
\begin{align}
    {\rm ZZ}&(\theta)e^{i \phi ({\rm X \otimes I})}{\rm ZZ}(\theta) = {\rm ZZ}(2\theta) \label{seq:strprop}\\
    &\times\exp [i\phi(\cos(\theta\pi/2)({\rm X \otimes I})+\sin(\theta\pi/2)({\rm Y \otimes Z}))],\nonumber
\end{align}
where $\theta$ is proportional to $J_{\rm Z}$ and the remaining ${\rm ZZ}(2\theta)$ rotation can, in principle, be removed by setting $\theta{\to}{-}\theta$ in one of the two initial ${\rm ZZ}(\theta)$s. This operation can be understood as a ${\rm ZZ}(\theta)$ unitary growing the ${\rm X}$ operator into a ${\rm Y}{\otimes}{\rm Z}$ one, and similar relations hold for all basis combinations. Moreover, Eq.~\eqref{seq:strprop} governs the growth of general strings, as ${\rm X}$ can denote a particular site on a larger string operator. The special case of $\theta{=}1$, corresponding to the CZ gate, results in a complete propagation.

Consider a hopping operator in a single direction, for example along an XX link. In particular, take four qubits, labeled 1,2,3, and 4, arranged such that the pairs (1,2) and (3,4) form ZZ links and qubits (2,3) are connected by an XX link. After applying two cycles of $U_F$ (with discarded YY terms), we can instead interpret it as the first hopping along XX followed by the second one that is propagated according to Eq.~\eqref{seq:strprop}. In the average Hamiltonian, for large angle $\theta$, this effectively realizes a ${\rm X}_2{\rm X}_3+{\rm Z}_1{\rm Y}_2{\rm Y}_3{\rm Z}_4$ operator which corresponds to complex-fermion dynamics (Extended Data Fig.~\ref{fig:ED_FermionEncoding}e). Note that ${\rm X}_2{\rm X}_3$ and ${\rm Z}_1{\rm Y}_2{\rm Y}_3{\rm Z}_4$ commute so there are no Trotter errors, but in principle there can be additional terms from other sites. In ED Fig.~\ref{fig:ED_FermionHopping}b, we numerically evaluate the particle conservation of the effective depth-6 Hamiltonian (without the $2\theta$ term in Eq.~\eqref{seq:strprop}) and see that indeed the particle creation is suppressed most when ${\rm ZZ}(\theta)$ realizes a CZ gate.

\naturepar{Fermion hopping and density-density correlations}
In our quench experiments with fermionic Hamiltonians, we study time dynamics of two initialized particles (Fig.~\ref{fig:fig4}). While the density of complex fermions at each step (Extended Data Fig.~\ref{fig:ED_FermionHopping}a) can reveal many spatial features, transport properties need to be inferred from other observables. For example, we could measure longer Pauli strings that include hopping terms (Extended Data Fig.~\ref{fig:ED_FermionEncoding}e). Alternatively, density-density correlations,
\begin{equation}
    G_{ij} = \braket{n_i n_j}-\braket{n_i}\braket{n_j},\label{seq:Gij}
\end{equation}
can also capture transport behavior. 
Intuitively, if a particle starts at site A and moves to site B, the density at sites A and B should be anticorrelated due to the particle leaving site A in order to appear at B.
In Fig.~\ref{fig:fig4}e, we plot a horizontal cut of $G_{ij}/\braket{n_i}$, which is additionally normalized by the density of the reference site $i$, and in Extended Data Fig.~\ref{fig:ED_FermionHopping}b we present it for a two-dimensional neighborhood of the reference site.

Furthermore, for free-fermion states, we can show that the connected density-density correlations in Eq.~\eqref{seq:Gij} directly capture hopping strength. After expanding the density operators, $n_i{=}a^\dagger_i a_i$, the correlation function becomes
\begin{equation*}
    G_{ij} = \braket{a^\dagger_i a_i a^\dagger_j a_j}-\braket{a^\dagger_i a_i}\braket{a^\dagger_j a_j},
\end{equation*}
in terms of complex-fermion operators. For free-fermion states, the four-body term can be further decomposed into two-body terms through Wick's theorem~\cite{wick_evaluation_1950}, 
\begin{align*}
    \braket{a^\dagger_i a_i a^\dagger_j a_j}=&\braket{a^\dagger_i a_i}\braket{a^\dagger_j a_j}-\braket{a^\dagger_i a_j}\braket{a^\dagger_j a_i}\\ &+\braket{a^\dagger_ia^\dagger_j}\braket{a_ja_i},
\end{align*}
where the last term vanishes when the particle number is conserved. This gives the expression for connected density-density correlations,
\begin{equation*}
    G_{ij} = -\abs{\braket{a^\dagger_j a_i}}^2+\abs{\braket{a^\dagger_i a^\dagger_j}}^2,
\end{equation*}
which contains the negative magnitude of the hopping current and a positive contribution from pair creation. Thus, for free-fermion states with well-defined particle number, such correlations should be negative, with a magnitude proportional to squared hopping strength.

\naturepar{Fermion exchange protocol}
Here we describe elements of the fermion exchange experiment presented in Fig.~\ref{fig:fig4}f. These experiments focus on four complex fermion sites embedded in the full experimental array, as depicted in Extended Data Fig.~\ref{fig:ED_Exchange}a. In order to perform exact evolution without disturbing the rest of the system, we perform local gates by moving a subset of the atoms not involved in the exchange protocol to the storage zone. With this method, we can conveniently perform the required local gates without disturbing other atoms with either the Rydberg excitation or moving AOD traps.

The full sequence for the fermion exchange protocol is shown in Extended Data Fig.~\ref{fig:ED_Exchange}b-c. In the first step, we use a sequence of three two-qubit unitaries to realize the three-body interaction term $R$ which creates the superposition of zero and two fermions at sites A and D. Then, we apply two different hopping steps, along YY and XX links (Extended Data Fig.~\ref{fig:ED_FermionEncoding}d-e). Since the Majorana hopping terms along  XX and YY links do not conserve the particle number, we realize particle-conserving hopping through a sequence of four parallel two-qubit gate operations (Extended Data Fig.~\ref{fig:ED_Exchange}b-c). Finally, we apply $R$ a second time to read out the exchange phase. In the absence of hopping, this final gate would complete the creation of two fermions at sites A and D, but due to the -1 exchange phase, we expect zero fermions in the final image (if no errors are present).

We benchmark our ability to realize particle-conserving complex fermion hopping terms by deterministically creating either 0 or 2 fermions, as shown in Extended Data Fig.~\ref{fig:ED_Exchange}d. To create two fermions, here we apply $R$ twice in a row (an alternative method to the local single-qubit gates used in Fig.~\ref{fig:fig4}a which also benchmarks our $R$ unitary). We observe that applying the hopping unitary preserves the vacuum state and hops the two fermions at sites A and D to sites B and C, as expected. 
\new{The results for both the full exchange and the control protocols match expectations and improve with postselection, providing evidence for observation of the exchange phase. A more robust protocol could involve measuring the full Ramsey fringe, for example by applying a variable phase to the two-fermion basis state before the final partial-creation operator.} Extended Data Fig.~\ref{fig:ED_Exchange}e shows data for the full exchange experiment, including some of the intermediate steps not shown in Fig.~\ref{fig:fig4}f.

\naturepar{Fermi-Hubbard implementation}
Here we provide more information about the implementation of Fermi-Hubbard quantum simulations in Fig.~\ref{fig:fig5}. To prepare the initial checkerboard configuration, we apply local gates on 16 of the sites after state preparation (Extended Data Fig.~\ref{fig:ED_FermiHubbard}a). For this circuit, we start with all operators on ZZ links with -1 value (by flipping the state of one of the data qubit sublattices after the CY operation). Then we use local Raman gates to flip the required ZZ operators to prepare the correct pattern of localized fermions. Here we use local \deleted{Raman} \new{single-qubit} Z gates and convert them to local X and Y gates using global $\pi$/2 Raman pulses. These local gates also flip four columns of plaquettes, which we pre-compensate for by flipping ancilla measurement results in the decoder.

We decompose each Floquet step of our simulations into (a) fermion hopping within each half (b) coupling between the two halves to realize the spin interaction term. For independent hopping within each half of the system, we turn off the gates along XX and YY links connecting the two halves by modifying the atom motion pattern (Extended Data Fig.~\ref{fig:ED_FermiHubbard}b). The spin interaction gates commute with the final measurement Z basis, so for the data in Fig.~\ref{fig:fig5}e-f we omit the final interaction step (e.g., after the first Floquet round both circuits are exactly the same data).

Concretely, through the entangling operations between the two halves, we realize contact fermion interactions with four-body terms $({\rm Z}_{\uparrow} {\rm Z}_{\uparrow}){\times}({\rm Z}_{\downarrow} {\rm Z}_{\downarrow})$ across the two halves. In terms of Majorana operators, these terms are proportional to $(ic_ic_{i'})_{\uparrow}(ic_{i}c_{i'})_{\downarrow}$; in terms of complex fermions, they are of the form $4n_{\downarrow}n_{\uparrow}{-}2n_{\downarrow}{-}2n_{\uparrow}$. The first-order Floquet Hamiltonian is therefore given by,
\begin{equation}
    H = \sum_{\sigma\in{\uparrow,\downarrow}}\sum_{\braket{i,j}}c_{\sigma,i}c_{\sigma,j} + U(c_ic_j)_{\uparrow}(c_ic_j)_{\downarrow},
\end{equation}
which reproduces the usual Fermi-Hubbard model, Eq.~\eqref{eq:fh}, when the complex fermion particle number is conserved.

\naturepar{Data Availability}
The data that supports the findings of this study are available from the corresponding author on \deleted{reasonable} request. \\

\noindent\textbf{Acknowledgments}
We thank A. Young for insights about particle statistics and feedback on the manuscript, and G. Semeghini for useful discussions and feedback on the manuscript. We further thank M. Aidelsburger, G. Baranes, J. P. Bonilla Ataides, J. Feldmeier, V. Menon, A. Stern, and N. Uğur Köylüoğlu for helpful discussions. We gratefully acknowledge L. Zheng, P. Sales Rodriguez, D. Paquette, and T. Karolyshyn for support with FPGA electronics. All experiments were performed on a machine built and operated at Harvard. We acknowledge financial support from the US Department of
Energy (DOE Quantum Systems Accelerator Center, grant
number DE-AC02-05CH11231\new{, and the QUACQ program, grant number DE-SC0025572}), the
DARPA ONISQ program (grant number W911NF2010021),
the DARPA IMPAQT program (grant number HR0011-23-3-0030),
the DARPA MeasQuIT program (grant number HR0011-24-9-0359), the
Center for Ultracold Atoms (an NSF Physics Frontiers
Center), the National Science Foundation (grant numbers PHY-2012023 and CCF-2313084), IARPA and the
Army Research Office, under the Entangled Logical Qubits
program (Cooperative Agreement Number W911NF-
23-2-0219), Wellcome Leap Foundation under the Quantum for Bio program, and QuEra Computing. S.J.E. acknowledges support from the National Defense Science and Engineering Graduate (NDSEG) fellowship. T.M. acknowledges support from the Harvard Quantum Initiative Postdoctoral Fellowship in Science and Engineering. D.B. acknowledges support from the NSF Graduate Research Fellowship Program (grant DGE1745303) and the Fannie and John Hertz Foundation. N.M. acknowledges support by the Department of Energy Computational Science Graduate Fellowship under award number DE-SC0021110. S.S. acknowledges support by U.S.
National Science Foundation grant DMR-2245246.
\\

\noindent\textbf{Author contributions} 
S.J.E., M.K., A.A.G., T.M., D.B., S.H.L., H.Z., S.E., and M.X. contributed to the building of the experimental setup, performed the measurements, and analyzed the data. M.K., N.M., M.C., and S.O.  contributed to theoretical analyses and understanding of results. M.K. and N.M. contributed to the theoretical development of fermion simulation and Chern number measurement schemes. J.C. developed FPGA electronics. All work was supervised by S.F.Y., S.S., M.G., V.V., and M.D.L. All authors discussed the results and contributed to the manuscript.
\\

\noindent\textbf{Competing interests:} M.G., V.V., and M.D.L. are co-founders\new{, M.G. V.V., M.D.L, and H.Z. are shareholders, S.F.Y.’s spouse is a co-founder and shareholder}, V.V. is \new{Chief
Technology Officer, M.D.L. is Chief Scientist}, and H.Z., J.C., and S.O. are employees of QuEra Computing.\\

\noindent\textbf{Correspondence and requests for materials} should be addressed to M.D.L.\\

\setcounter{figure}{0}
\newcounter{EDfig}
\renewcommand{\figurename}{Extended Data Fig.}

\begin{figure*}
\includegraphics[width=18.0cm]{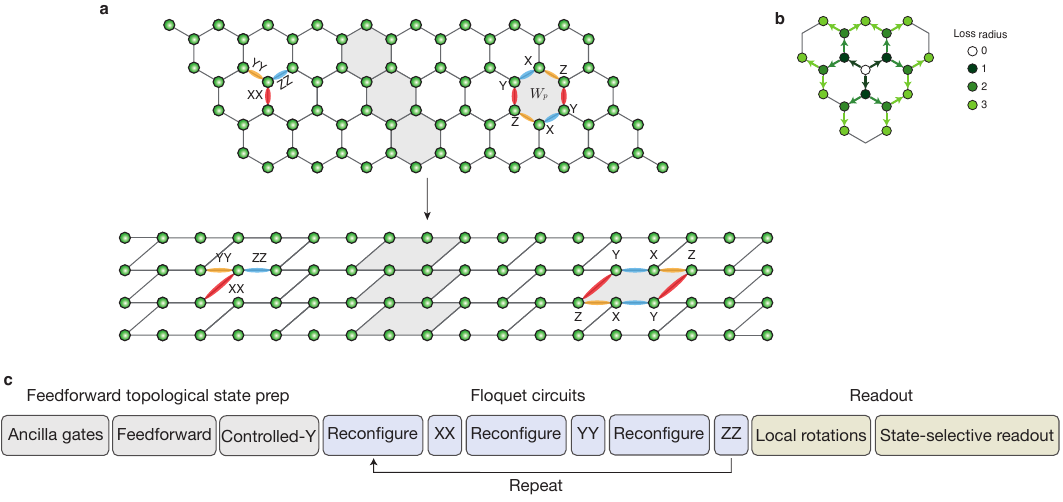}
\caption{\textbf{Honeycomb lattice layout and experiment sequence.} \textbf{a,} Mapping from honeycomb geometry into experimental array geometry used, indicating link orientations, an example column, and an example plaquette orientation. We use this mapping in order to shrink the number of rows needed, as well as put the atoms into a rectangular grid which is convenient for atom motion and local Raman operations. \textbf{b,} Loss radius definition for error detection based on atom loss. The central (white) atom is the reference point, and for loss radius of 0, we postselect on this atom being present at the end of the circuit. For a larger loss radius, we postselect on atoms within a certain local region being present. For string observables, we perform this procedure for all atoms within the string. \textbf{c,} High-level overview of the experimental sequence used in these experiments, including feedforward topological state preparation, Floquet evolution, and measurement steps. The Fermi-Hubbard quantum simulations in Fig.~\ref{fig:fig5} have an additional part of the Floquet circuits for engineering the onsite density-density interactions.
}
\refstepcounter{EDfig}\label{fig:ED_ExperimentSequence}
\end{figure*}

\begin{figure*}
\includegraphics[width=18.0cm]{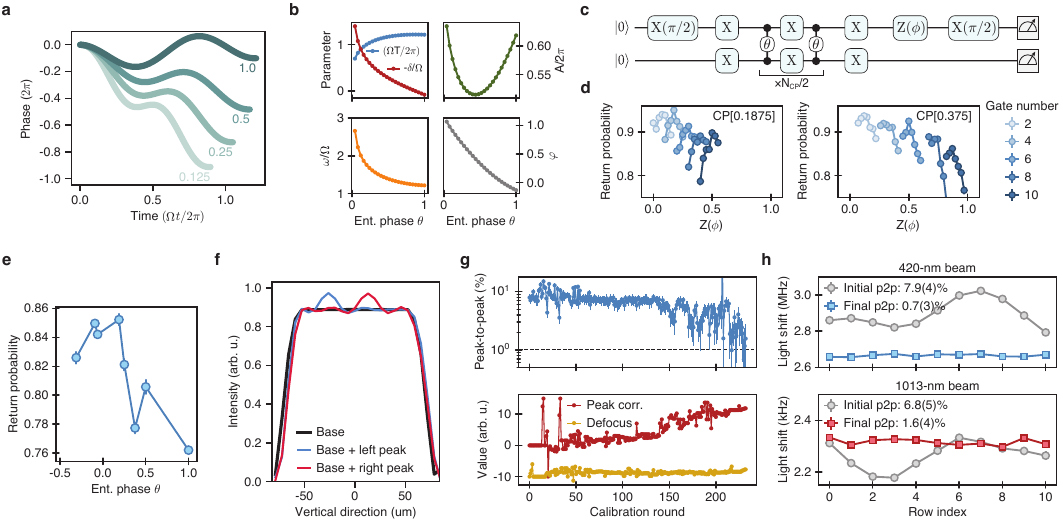}
\caption{\textbf{Tunable CPHASE gates and automated calibration procedures.} \textbf{a,} 
Example entangling gate phase profiles for different entangling phases, with detuning included as a linear phase term. As the entangling phase becomes larger, the gate becomes longer. An entangling phase of $\theta{=}1$ is equivalent to the CZ gate profile. \textbf{b,} Theoretical parameters for different entangling phase gates, where the gate has a constant amplitude profile and a phase profile given by the cosine function $A\cos(\omega t{+}\varphi)$~\cite{evered_high-fidelity_2023}. We optimize these gates experimentally with parameter scans in the vicinity of theoretical values~\cite{evered_high-fidelity_2023}. \textbf{c,} Circuit used to benchmark and calibrate the entangling phase gates, adapted from the approach in~\cite{mi_timecrystal_2022}. One atom is prepared in $\ket{0}$ and the other in $\ket{+}_y {=}(\ket{0}{+}i\ket{1})/\sqrt{2}$ with a local Raman gate. Then a series of CPHASE gates are applied to the atom pair, which effectively rotates the phase of the initial $\ket{+}_y$ state. A precalculated final ${\rm Z}(\phi)$ gate ensures that both atoms return to $\ket{0}$ in the absence of errors. Errors, for example coming from a miscalibrated entangling phase or qubit loss/leakage during the gate, will reduce the probability of finding both atoms in $\ket{0}$ at the end of the circuit (the return probability). \textbf{d,} Data used to extract the entangling phase in Fig.~\ref{fig:fig1}d, shown for two different CPHASE gates. For different numbers of gates applied, we scan the ${\rm Z}(\phi)$ gate before the final local $\pi$/2 pulse in order to extract the amount of phase accumulated. The trend shows how much phase has been accumulated over the 10 gates, and the reduction in the peak return probability is a result of gate errors. Note that for a smaller number of gates, we use the same Raman pulse sequence as for the 10 gate sequence and only reduce the number of CPHASE gates. \textbf{e,} Comparison of return probability after 20 CPHASE gates with this benchmarking method for different entangling phases, with the point $\theta{=}1$ being the CZ gate. We attribute the non-monotonic behavior to varying levels of calibration between the gates. These values can be compared to a return probability after 0 gates of 0.952(3) owing to errors from components separate from the CPHASE gates. We note that this calibration sequence is not a proper measure of fidelity, and we rather use this measurement to compare the different CPHASE gates to each other. The CZ gate was benchmarked right after taking this data with a fidelity of 99.4\% in the global RB sequence utilized in~\cite{evered_high-fidelity_2023} (the slightly lower fidelity than our typical 99.5\% operation can be attributed to the increased scattering from the closer intermediate-state detuning used here). \textbf{f,} Example calculated Rydberg beam profile and the effect of adding two separate peak-correction holograms. The local peak corrections are added to the base holograms with variable weights on the Rydberg beam SLMs, in addition to correcting for more global Zernike aberrations. \textbf{g,} Example automatic calibration routine showing the homogenization of our Rydberg beam across an array. (top) The peak-to-peak variations in row intensity across the array during calibration, as measured by the differential light shift on the hyperfine qubit. (bottom) Example of how the defocus and one of the peak corrections change during this automated calibration procedure. \textbf{h,} Examples of the measured light shift across the rows of our atom array for the two Rydberg beams, comparing the uniformity before and after the automated calibration procedure. We also ensure that the columns are uniform, although they are naturally more homogeneous due to the beam geometry.}
\refstepcounter{EDfig}\label{fig:ED_EntanglingGates}
\end{figure*}

\begin{figure*}
\includegraphics[width=18.0cm]{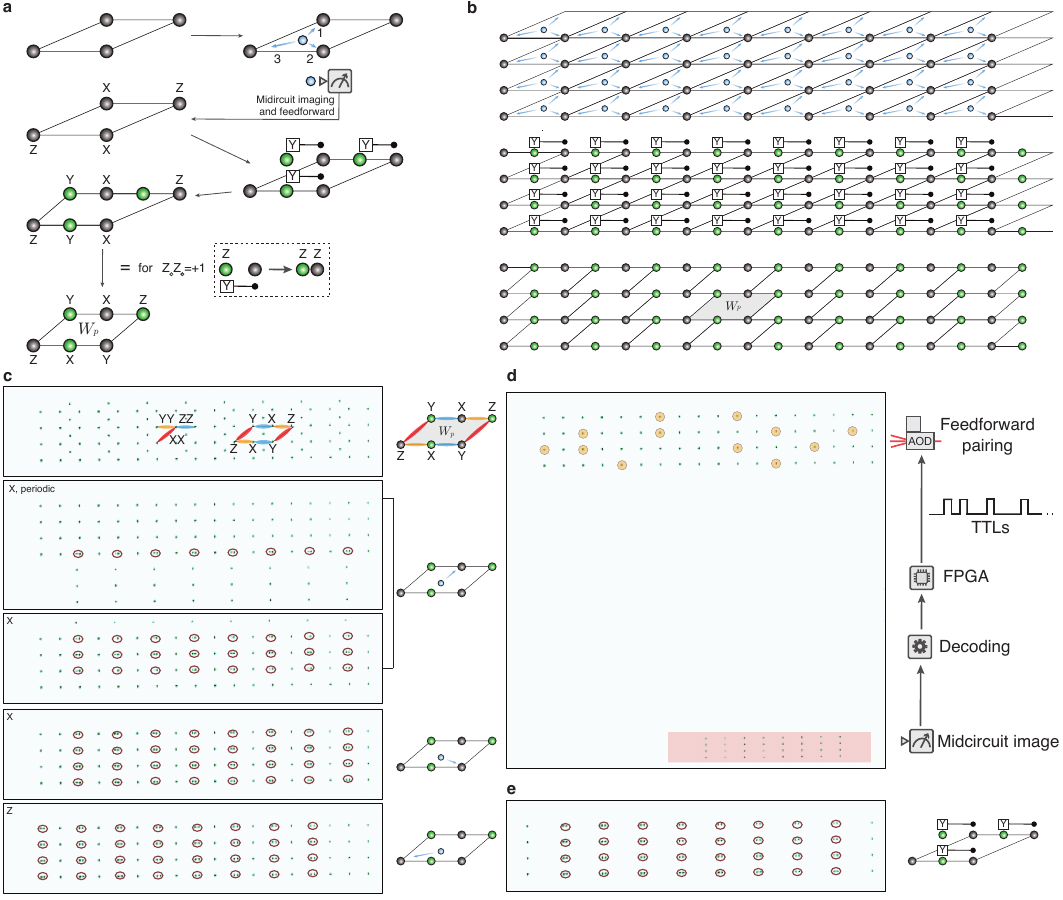}
\caption{\textbf{Measurement-based topological state preparation.} 
\textbf{a,} Experimental protocol used for measurement-based preparation of the topological state. First, a ZXXZ toric code state is prepared on one data qubit sublattice (black circles) using ancilla qubits (blue circles). After midcircuit measurement and feedfoward, this prepares the ZXXZ toric code state. The weight-four ZXXZ operators are then extended to weight-6 operators with parallel controlled-Y (CY) operations with a fresh data qubit sublattice originally in $\ket{0}$ (green circles). In the case that the ZZ link (where the CY gates were applied) are +1, this weight-6 operator is equal to the plaquette operators $W_p$. This sequence indeed ensures that these ZZ operators are all +1, owing to the propagation of the single-qubit Z stabilizer on the green qubit sublattice through the CY operation, which results in the weight-2 ZZ stabilizers. 
\textbf{b,} Implementation of the encoding steps in parallel across the full array. The top row of ancillas also performs gates with the bottom row of data qubits.
\textbf{c,} Experimental layout, with gate orientation and an example plaquette shown, and ancilla qubit motions for preparing the ZXXZ surface code state on one half of the data qubits. We perform the periodic ancilla measurement step first, such that all other qubits are still in $\ket{0}$ and do not experience most of the Rydberg gate errors.
\textbf{d,} Atom positions for midcircuit readout and feedfoward, with an example feedforward pattern applied (indicated by orange circles). The midcircuit image is analyzed, and then based on our decoding algorithm, an FPGA outputs a series of digital TTLs which determine which atoms have a local Raman gate applied.
\textbf{e,} Data qubit positions for the parallel controlled-Y (CY) operation, completing the state preparation sequence. The sublattice denoted by black circles gets picked up by AOD traps and shifted to the left to perform the parallel CY. This sublattice then remains in AOD traps for the Floquet circuit.
}
\refstepcounter{EDfig}\label{fig:ED_AncillaMeasurement}
\end{figure*}

\begin{figure*}
\includegraphics[width=18.0cm]{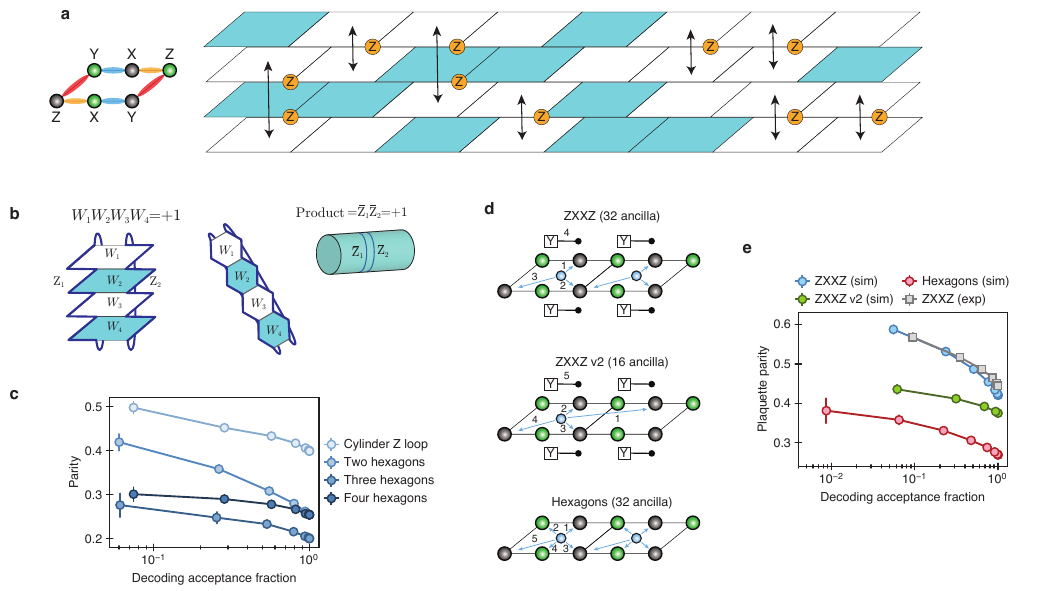}
\caption{\textbf{Midcircuit decoding and error detection method.} 
\textbf{a,} Example of decoding and feedforward pairing based on ancilla results. White (blue) plaquettes have the value of -1 (+1). After the pairing procedure, all plaquettes are prepared with +1 value in the absence of errors. Note that the local Z gates flip plaquettes within the column to perform the 1D matching, but do not flip plaquettes in adjacent columns.
\textbf{b,} Illustration of decoding postselection method based on parity of ancilla measurement results within a column, owing to the cylindrical geometry used in this work. The product of plaquettes within a column is equal to two Z loops enclosing the cylinder, which must be +1 in the absence of errors.
\textbf{c,} Larger loops measured for the initial state, as a function of decoding postselection (the maximum decoding postselection is plotted in Fig.~\ref{fig:fig2}\deleted{d}\new{c}). We attribute the fact that the loop enclosing four hexagons is higher parity than the loop enclosing three hexagons due to certain differences in the measurement, for example the loops enclosing the cylinder are measured in the Z basis which doesn't require local Raman gates to rotate individual qubits.
\textbf{d,} Comparison of different state preparation methods. The method we use in this work is based on the ZXXZ toric code prepared with 32 ancilla qubits. This method can also be adapted to require only 16 ancilla qubits. A third method (``Hexagons") consists of directly measuring the weight-6 plaquette operators with 32 ancilla qubits, requiring a slightly deeper state preparation circuit. \textbf{e,} Numerical simulation of plaquette parity for these different state preparation methods, for the \new{phenomenological} noise model which matches our experimental data \new{(depolarizing noise of strength 0.027 on all qubits after every two-qubit gate layer)}. The method that we use in this work (ZXXZ) performs the best in numerics, both for no postselection as well as with decoding postselection.
}
\refstepcounter{EDfig}\label{fig:ED_FeedforwardMethods}
\end{figure*}

\begin{figure*}
\includegraphics[width=18.0cm]{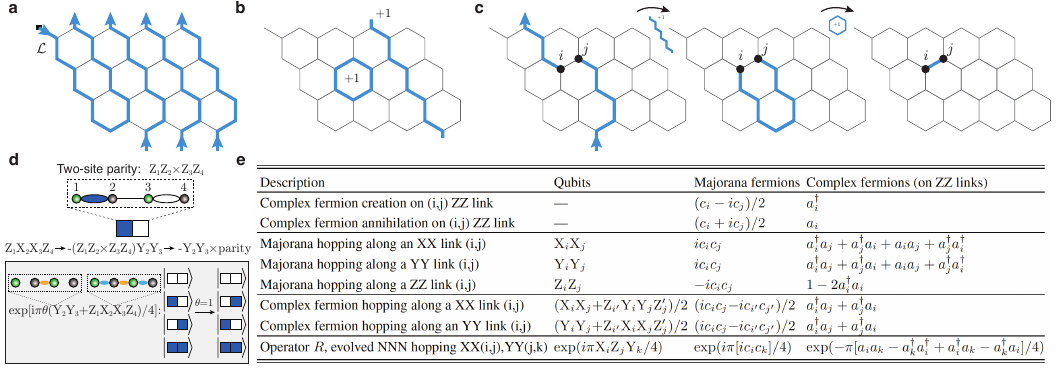}
\caption{\textbf{Fermion encoding description.} Details of the fermion-to-qubit mapping and its origins. \textbf{a,} An example Jordan-Wigner (JW) ordering spanning a honeycomb lattice on a cylinder. In our case the string is constructed as a product of link operators (see text). \textbf{b,} The operators which are fixed to +1 (stabilizers) on the topological state include the hexagonal plaquettes and the loops around the cylinder. \textbf{c,} A local fermion hopping term between two sites results in a macroscopic JW operator. The underlying long-range entanglement enables reducing this operator to a local one by multiplying the string with the relevant stabilizers.
\textbf{d,} Construction of complex-fermion hopping in terms of spin operators. The length-4 Pauli string cancels the particle-creation terms resulting from the length-2 hopping operator. \textbf{e,} Table describing operator mapping between qubits and fermions. Sites $i$,$k$ belong to the even sublattice and $j$ belongs to the odd sublattice. Sites $i'$,$j'$ complement $i$,$j$ along the ZZ links.}
\refstepcounter{EDfig}\label{fig:ED_FermionEncoding}
\end{figure*}

\begin{figure*}
\includegraphics[width=18.0cm]{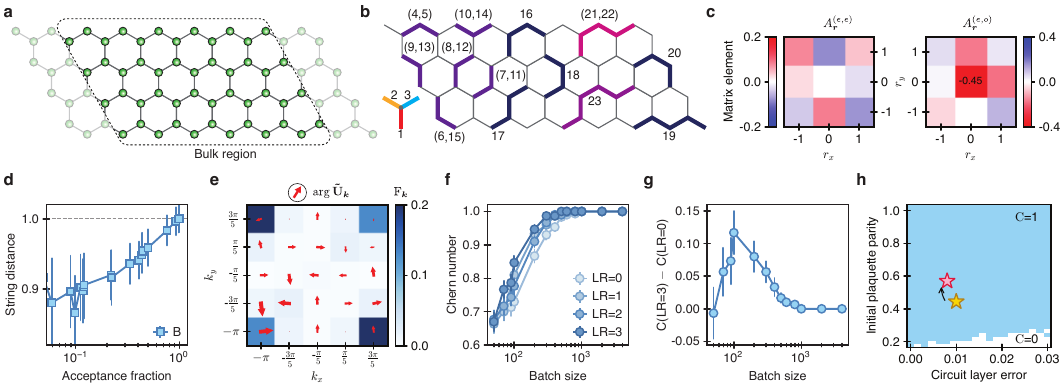}
\caption{\textbf{Chern number analysis.} 
\textbf{a,} The central 12 columns are chosen as the bulk region of the system.  \textbf{b,} The measured open Pauli strings, numbered according to ordering on the horizontal axis of Fig.~\ref{fig:fig3}c.  \textbf{c,} Majorana correlations in the unit-cell basis, obtained from the open strings. The correlation matrix is truncated at distance $\pm$1 (length-6 strings) to avoid system boundaries when starting from a reference site in the bulk. The $(o,e),(o,o)$ matrices are evaluated similarly, and related to $(e,o),(e,e)$, respectively, through symmetry relations.  \textbf{d,} The total variation distance of the measured string distribution (in phase B) from the ideal theory values with increasing error detection. The value is normalized by the point with least postselection and improves by $\sim$10\% for optimal error detection. \textbf{e,} Discretized Berry curvature ${\rm F}_{\bm k}$ whose sum over the 1st Brillouin zone is the Chern number. The red arrows represent the discretized phase potential obtained from normalized eigenstate overlaps, with the arrow vector given by $\arg\tilde{\rm U}_{\bm k}{=}(\arg {\rm U}^{\hat{x}}_{\bm k},\arg {\rm U}^{\hat{y}}_{\bm k})$. \textbf{f,} The Chern number for different values of loss radius (LR) postselection, evaluated through bootstrapping with 300 trials as a function of a single batch size. The samples are drawn uniformly from the data set with replacement, and the samples from strings identified with each other are afterwards further averaged. \textbf{g,} Change in the Chern number between postselection at loss radius 3 and 0, as a function of batch size. \textbf{h,} Phase diagram of the Chern number as a function of per-layer error and initial plaquette parity values, obtained through noisy numerical simulations. The stars denote points corresponding to the phenomenological noise model inferred from experimental measurements (red with maximal postselection).
}
\refstepcounter{EDfig}\label{fig:ED_MajoranaChern}
\end{figure*}

\begin{figure*}
\includegraphics[width=18.0cm]{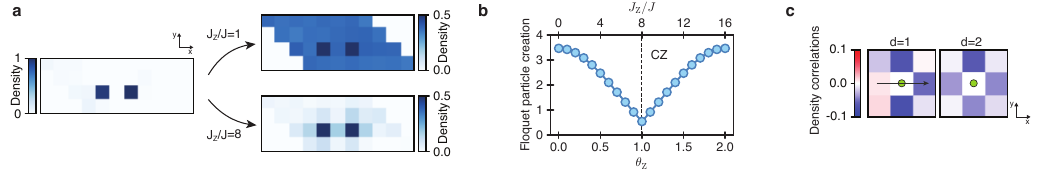}
\caption{\textbf{Fermion hopping dynamics.} \textbf{a,} Spatial distribution of complex-fermion density, $n_i$, after evolution for depth-11 from the initial two-fermion state. The density values are obtained as $n_i{=}(1{-}{\rm Z}_{i}{\rm Z}_{i'}/\bar{zz})/2$, where $\bar{zz}$ is the background magnitude of the ZZ link operators in the initial state, $\bar{zz}{=}0.728(3)$. \textbf{b,} The particle creation of the effective depth-6 Floquet Hamiltonian $H_{F,6}$ obtained from two rounds of the basic Floquet unitary in Eq.~\eqref{seq:Uf} with the evolution time of individual unitaries corresponding to that of the quench experiment, $Jt{=}0.125$. The particle creation is quantified here as a matrix 2-norm of the commutator between the bulk Floquet Hamiltonian and the particle number operator, $\lvert\lvert[H_{F,6},N_{\rm tot}]\rvert\rvert_2$. \textbf{c,} A spatial plot of density-density correlations normalized by the density of the reference site (green dot), $G_{ij}\,{/}\braket{n_i}$, in a small neighborhood of the reference site. The arrow represents the orientation of the one-dimensional cross-sections plotted in Fig.~\ref{fig:fig4}e.
}
\refstepcounter{EDfig}\label{fig:ED_FermionHopping}
\end{figure*}

\begin{figure*}
\includegraphics[width=18.0cm]{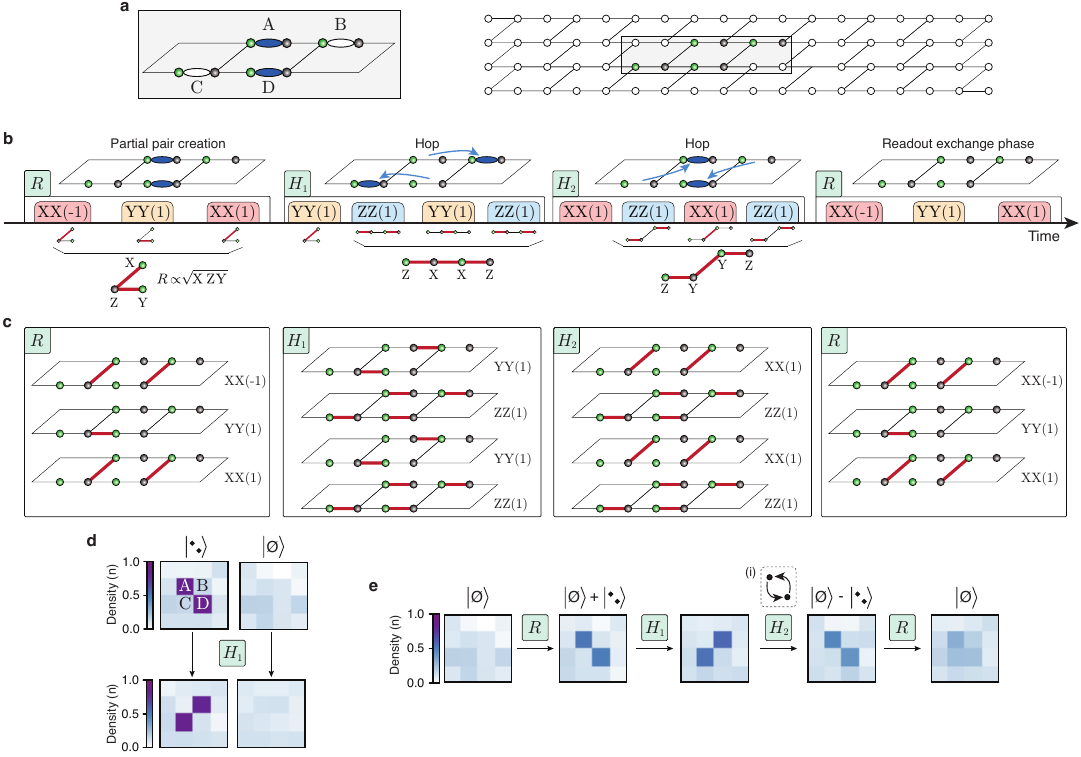}
\caption{\textbf{Fermion exchange protocol.} \textbf{a,} Four complex-fermion sites involved in the fermion exchange experiments, and where they are embedded in the full lattice. \textbf{b,} Full gate sequence for the exchange protocol (partial pair creation, first hop, second hop, and readout of the exchange phase by repeating the partial pair creation). Four-body interaction terms for the hopping sections are realized by a sequence of two-qubit gate applications. \textbf{c,} Depiction of qubits undergoing the two-qubit gates at each time step, denoted by red lines. The black qubits are the atoms that are picked up by AOD traps to perform the gates, and the layout is designed such that gates can be applied in parallel at each time step, as illustrated in the diagrams. \textbf{d,} Demonstration of the hopping gate H$_1$ for the two-fermion basis state (created by applying $R$ twice) and the fermionic vacuum state. \textbf{e,} Additional data for the exchange protocol, showing additional intermediate steps of the process. During the middle three steps, the state of the system is described by a superposition of zero fermions and two fermions at different positions in the lattice.
}
\refstepcounter{EDfig}\label{fig:ED_Exchange}
\end{figure*}

\begin{figure*}
\includegraphics[width=18.0cm]{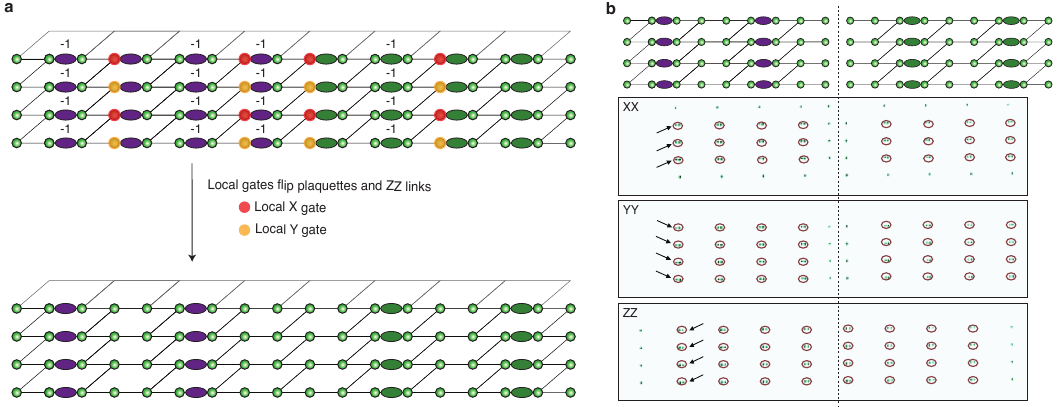}
\caption{\textbf{Fermi-Hubbard implementation.} \textbf{a,} Local gates used to prepare the checkerboard pattern in Fig.~\ref{fig:fig5}d. In order to correct the plaquettes, we pre-compensate with the decoding algorithm such that local gates both flip the plaquettes to all be +1 and also initialize the operators on ZZ-links to give the correct complex fermion configuration. \textbf{b,} Depiction of how the system is cut in half for the Fermi-Hubbard implementation. Atom moves are shown for the XX, YY, and ZZ terms (no periodic boundary conditions are used here unlike in earlier experiments). These gates comprise the rest of the Floquet circuit in addition to the interaction terms shown in Fig.~\ref{fig:fig5}c. Note that the middle links are turned off for the XX and YY gates to realize independent hopping on the two halves.
}
\refstepcounter{EDfig}\label{fig:ED_FermiHubbard}
\end{figure*}

\end{document}